\documentclass[preprint,5p,number]{elsarticle}
\usepackage[T1]{fontenc}
\usepackage[english]{babel}
\usepackage{amsmath}
\usepackage{amsfonts}
\usepackage{amssymb}
\usepackage{latexsym}
\usepackage{graphics}
\usepackage{graphicx}
\usepackage{hyperref}
\usepackage{epigraph}
\usepackage{epsf}
\usepackage{aniszewski,color}
\title{A New Approach to Sub-grid Surface Tension for LES of Two-phase Flows}

\author[coria]{W. Aniszewski\corref{cor1}}  
\ead{aniszewski@coria.fr}

\cortext[cor1]{Corresponding Author. email: aniszewski@coria.fr; tel. +33 (0)2 32 95 36 73,  CNRS-UMR6614 CORIA, Universite de Rouen, Site Universitaire du Madrillet - BP 12, 76801 Saint Etienne du Rouvray cedex}

\author[imc]{A. Bogus\l awski}

\author[imc]{M. Marek}

\author[imc]{A. Tyliszczak}

\address[coria]{CNRS UMR 6614 - CORIA Rouen, Site Universitaire du Madrillet, Saint Etienne du Rouvray, France}
\address[imc]{Institute of Thermal Machinery, Czestochowa University of Technology, Czestochowa, Poland}

\journal{Journal of Computational Physics}

\begin{document}
\bibliographystyle{plain}
    \begin{abstract}
In two-phase flow, the presence of inter-phasal surface -- the interface -- causes additional terms to appear in LES formulation. Those terms were ignored in contemporary works, for the lack of model and  because the authors expected them to be of negligible influence.  However, it has been recently shown by \textit{a priori} DNS simulations that the negligibility assumption can be challenged. In the present work, a model for one of the sub-grid two-phase specific terms is proposed, using deconvolution of the velocity field and advection of the interface using that field. Using the model, the term can be included into LES. A brief presentation of the model is followed by numerical tests that assess the model's performance by comparison with \textit{a priori} DNS results. 
    \end{abstract}


    \maketitle

\section{Introduction}

It is our focus in this work to present an algorithm that allows for calculation of sub-grid surface tension term that appears in Navier -- Stokes equations within Large Eddy Simulation (LES) formulation when two-phase flow is considered. To make the required calculations, an entire numerical setup involving a solver, two-phase flow modules and a sub-grid model have been prepared. In the article  solving Navier -- Stokes equations, the Volume of Fluid (VOF) method and the  Level Set (LS) method  will be described only in minimum, providing the reader with necessary literature, while the focus is on the implementation of the Coupled Level Set and Volume of Fluid method and the ADM-$\tau$ model of sub-grid surface tension, which we describe in most detailed manner.  Similarly, the introduction is focused solely on reviewing Large Eddy Simulation of two-phase flow and not two-phase flows in general. 

Until the 1990s, simulations of turbulent flows involved mainly one-phase (gaseous) flows; behaviour of second phase was, e.g. in work of Elghobashi (1984)  \cite{elgho}, calculated using discrete particles (undeformable and much smaller than mesh size) so that their concentrations could be predicted  \cite{elgho}. Similar work was published by Eaton in 1994  \cite{eaton}. 

Explicit interface modelling using methods such as VOF was not applied in LES until the 2000s. Conference paper of Alajbegovic gave first accounts of applying  Large Eddy Simulations to multiphase flow  \cite{alajbegovic}, where he even proposed a closure for sub-grid scale surface tension,  but results were criticized as being weakly documented  \cite{shiriani}, and having little connection with changes of the interface topology \cite{liovic-conf}. In 2000s, many authors used standard LES implementations together with VOF two-phase advection scheme. Such approach uses filtered (large-scale, resolved) velocity field in which  advection, traced by VOF or similar method takes place, and all specific sub-grid terms resulting from the presence of the interface are ignored  \cite{macer}. In  \cite{lakehal2002}, Lakehal performed LES using Germano's dynamic procedure of bubbly shear flow and investigated bilateral dependence between turbulence parameters and bubbly phase. Klein \& Janicka  \cite{k-j}  simulated film breakup, and in 2004 a SAE paper of de Villers et al.  \cite{devillers} gave account of a jet breakup simulated with quasi-realistic conditions (with Reynolds number equal to $15000$), including an investigation of resolved droplet distributions. Similar simulation in 2007 by Bianchi et al.  \cite{bianchi} included more realistic generation of inlet conditions, turbulence spectrum analysis, finer grid and more realistic density ratio, with a little  lower Reynolds numbers. Still, interface-specific sub-grid contributions were not accounted for. Similar in character was the work of Menard  \cite{menard}, in which authors used CLSVOF technique to track the interface. It is important to stress that the authors of  \cite{menard,bianchi,devillers} all performed simulations in 3D, that were rather time-consuming (between one  \cite{devillers} and several  \cite{menard} months) despite being LES calculations. Many useful information about ligament formation, creation of larger droplets and their secondary breakup were thereby made available, however, care is advised in interpretation of such results as some of small scale structures ``cannot be trusted''  \cite{desjardins2010}.

We will now turn to works that directly precede some of the results presented in this paper. Articles by Labourasse et al.  \cite{labourasse}, Vincent et al.  \cite{vincent}, Toutant et al.  \cite{toutant} and Larocque et al.  \cite{larocque}, appearing between 2006 and 2010 were centered around careful examination of all sub-grid terms that appear in a two or multi-phase flow as a result of spatial filtering. Work by Trontin et al.  \cite{trontin} focused on creating a data base for development of two-phase SGS models by performing DNS of bubbly flow, including the question of subgrid contribution into the turbulent kinetic energy budget.  Labourasse et al.  \cite{labourasse} presented wide mathematical background for LES of two-phase flow, and defined respective tensors, among which $\tau_{rnn},$ the tensor connected with unresolved surface tension force, is modelled by present authors. All four articles were similar in concept, in that they were \textit{a priori} evaluations of LES sub-grid terms basing on DNS simulations. For example,  \cite{vincent} includes a phase-inversion problem similar to Rayleigh-Taylor instability, while  \cite{toutant} performs a simulation of a fluid droplet immersed in quasi-turbulent flow, similar in principle to Marek \& Tyliszczak  \cite{dropturb}. In all these cases, time evolution of all sub-grid tensors is investigated and magnitude of their components presented and classified in various ways -- so that it constitutes a valuable material for LES calculations. None of these works proposes any model yet.

\section{Description of the Flow}

We consider the one-fluid formulation for the incompressible two-phase flow  \cite{aris,labourasse,tsz}

\begin{equation}
\label{ns}
\frac{\partial \ub}{\partial t} + \nabla \cdot \ub\otimes\ub=\frac{1}{\rho}\left(\nabla\cdot\left(-p\ib+\mu\db\right) + \sigma\kappa\nb\delta_S \right) +\mathbf{f}_g,
\end{equation}
with continuity equation

\begin{equation}
\label{cont}
\nabla\cdot\ub=0.
\end{equation}
In (\ref{ns}), symbol $\db$ is used for  rate of strain tensor \[\db=\nabla\ub+\nabla^T\ub,\] while standard symbols $\ub,p,\mu,\kappa$ and $\nb$ are used respectively for velocity field, pressure, viscosity, interface curvature and interface normals. The $\mathbf{f}_g$ symbol stands for external (body) force such as gravity, while $\delta_S$ is the Dirac delta centered on the interface. The $\sigma\kappa\nb\delta_S$ is denoted $\mathbf{f}_s,$ by some authors  \cite{aniszewski, tsz}, and may be perceived as a surface tension force, which is of singular character, in that it is centered on the interface. The one-fluid formulation  \cite{kataoka}  is derived from formulations separate for every phase, such as (for $k$ phases) 

\begin{equation}
\label{4}
\frac{\partial \ub^k}{\partial t} + \nabla \cdot \ub^k\otimes\ub^k=\frac{1}{\rho^k}\nabla\cdot\left(-p^k\ib+\mu^k\db^k\right) +\mathbf{f}_g
\end{equation}
and

\begin{equation}
\label{cont1ph}
\nabla\cdot\ub^k=0
\end{equation}
with appropriate jump conditions applied on the interface, from which the $\mathbf{f}_s$ term of (\ref{ns}) originates.   In short, the jump conditions for incompressible flow with constant surface tension coefficient require jump conditions in the following form:

\begin{equation}
\label{jc}
\ub^1=\ub^2
\end{equation}
as the  continuity condition over the interface between phases $1$ and $2,$ and

\begin{equation}
\label{jc2}
\lbrack -p+2\mu\nb\cdot\db\cdot \nb \rbrack_s =  \sigma\kappa
\end{equation}
for Navier -- Stokes equation. In (\ref{jc2}) we use the jump notation $\lb x \rb_s=x^1-x^2.$

Paper by Labourasse  \cite{labourasse}, excellent coverage by Tryggvasson, Scardovelli and Zaleski  \cite{tsz} or one of present authors' PhD thesis  \cite{aniszewski} give  broad specification of jump conditions and one-fluid formulation for two-phase flow.

If we denote subdomains $\Omega_1,\Omega_2\subset\Omega$ as occupied by first and second phase, it needs  to be noted that interfacial surface $S$ may be introduced as a jump surface of one of the phases' characteristic function, for example

\begin{equation}
\label{chi}\chi^1(\xb)=
\begin{cases}
1 & \Leftrightarrow  \xb\in\Omega^1 \\
0 & \Leftrightarrow  \xb \notin\Omega^1 \vee x \in  S,
\end{cases}
\end{equation}
where $\Omega^1$ is a subdomain occupied by phase $1.$ Obviously, for two-phase flow we have $\chi^2=1-\chi^1.$ In such a situation, additional equation has to be formally considered with (\ref{ns}) and (\ref{cont}), namely transport of the phase indicator function, that is

\begin{equation}
\label{chitrans}
\frac{D\chi^1}{Dt}=0.
\end{equation}


\section{Numerical Approach}
\subsection{Large Eddy Simulation - Single-phase Flow}
\label{lesfiltrationsection}

The Large Eddy Simulation (LES)  \cite{smagorinsky,pope} bases conceptually upon \textit{spatial} filtering. Filtering is here defined as convolution with a chosen filter kernel. In one  dimension

\begin{equation}
\label{ff1}
\overline{u(x)}=\int G(x-x') \ub(x',t) dx'\mbox{ } \wedge \mbox{ } \int\limits_{-\infty}^{\infty}G(x)dx =1,
\end{equation}
where $G$ is the filter kernel. Multidimensional filtering is realized by a superposition of filters defined along three coordinate axes. 

The $G(x)$ is either only locally nonzero in physical space of $x\in\mathbb{R}$ or defined in the spectral space to filter out large wave numbers, i.e. its Fourier transform is zero almost everywhere.  

When single-phase flow is considered, the filtered Navier -- Stokes equations become:

\begin{equation}
\label{ff2}
\frac{\partial \overline{\ub}}{\partial t} + \nabla \cdot \underbrace{\overline{\ub\otimes\ub}}_I=\frac{1}{\rho}\nabla\cdot\left(-\overline{p}+\mu\overline{\db} \right) +\mathbf{f}_g
\end{equation}
with continuity equation for filtered velocity:

\begin{equation}
\label{ff3}
\nabla\cdot\overline{\ub}=0.
\end{equation}
 From the definition (\ref{ff1}), and examples of filters given below, it is clear that filtering is a linear operation; additionally, in (\ref{ff2}) commutation of filtering and differentiation is assumed.
In LES, by principle, only filtered variables are known, so $\ub$ field is unknown. Because of this, term $I$ in (\ref{ff2}) cannot be directly calculated and has to be closed, that is, expressed using only $\overline{\ub}.$  A symbol $\tau_{luu}$ is introduced in  \cite{labourasse} for this term, which is called sub-grid stress tensor. It has the following form: 

\begin{equation}
\label{ff4}
\tau_{luu}=\left(\overline{\ub\otimes\ub}-\overline{\ub}\otimes\overline{\ub}\right),
\end{equation}
which, substituted in (\ref{ff2}) yields 
\begin{equation}
\label{ff5}
\frac{\partial \overline{\ub}}{\partial t} + \nabla \cdot \left(\overline{\ub}\otimes\overline{\ub}+\tau_{luu}\right)=\frac{1}{\rho}\nabla\cdot\left(-\overline{p}+\mu\overline{\db} \right) +\mathbf{f}_g
\end{equation}
As it involves unknown non-filtered velocity field, number of closures exist for $\tau_{luu}.$ 

\subsection{Two-phase Flow}

In two-phase flow, as we remember by comparing respective forms of N-S equations (\ref{ff2}) and (\ref{ns}), at least one new term appears due to the presence of the surface tension force, that resulted from jump conditions being applied on the interface. The term is
\begin{equation}\label{efes}
\mathbf{f}_s=\sigma\kappa\nb\delta_S
\end{equation} 
and it undergoes filtering similarly to one-phase specific terms. Alternatively, as mentioned by Labourasse et al.  \cite{labourasse}, it is possible to perform the filtering on the left hand side of jump conditions (\ref{jc2}). In fact,  \cite{labourasse} proposes considering (\ref{jc2}) even in its more general form; thus,  instead of considering $\overline{\mathbf{f}_s}$ we would be considering the term

\[
 \sum\limits_k\overline{\left(\rho^k\ub^k\otimes(\mathbf{u_s}-\ub^k)-p^k\ib+\tb^k\right)\cdot\nb^k\delta_S},
\]

 where $\ub_s$ is the velocity of the interface. This strategy leads to severe complications, that is the need for closure not only for classical sub-grid term $\tau_{luu}$ but also amounts to formulating an approximate jump condition through the interface, therefore it is avoided  \cite{labourasse, vincent, toutant}. In this article, we thus adhere to the simpler approach, that is the filtering of (\ref{efes}), as described below. 
 
When considering surface-specific sub-grid (filtered) terms, Labourasse  \cite{labourasse} notes also that in general, filtering might not commute with surface differentiation\footnote{Although no numerical investigations pertaining to this issue exist at the moment.}, that is the use of operator $\nabla_s.$ This is a remark similar in nature to aforementioned issue of commutation between filtering and $\nabla.$ In our work, we do not introduce any corrections connected to this possible lack of commutability, since $\nabla\cdot\nb=\nabla_s\cdot\nb$ provided that $\nb$ extends off the interface \cite{tsz}. Hence, we will not introduce any new differentiation operators (in  \cite{labourasse}, new operator $\widehat{\nabla_s}$ is introduced in this context). 

When jump-conditions are introduced, phase-indicator functions $\chi^k,$ (\ref{chi}) are defined. These functions are filtered as well, as they are represented in any two-phase flow computation for example by levelset functions $\phi(x).$ They undergo advection under $\overline{\ub}$ velocity field described by equation 

\begin{equation}
\label{chiadv}
\frac{D\chi^k}{D t}=0,
\end{equation}
which is non-linear and due to filtering requires a closure. In contemporary literature this closure is ignored  \cite{toutant, vincent}. In the present work, while still not presenting any closure for this term (which we decided to call ``Sub-Grid Mass Transfer/Transport'', SMT), we give short remarks concerning it below.

Formal filtering of $\mathbf{f}_s$ should take the form

\begin{eqnarray}
\label{fsfilter}
\overline{\mathbf{f}_s} & = & \int\limits_\Omega G(\xb,\Delta)\mathbf{f}_s(\xb)d\xb \nonumber \\
               & = & \sigma\int\limits_\Omega G(\xb,\Delta)\left\lb(\nabla_s\cdot\nb(x))\nb(x)\delta_S\right\rb dx\\
               & = & \overline{(\nabla_s\cdot\nb)\nb\delta_S},
\end{eqnarray}
which is not computable, since $\nb$ is unknown in LES in favor of $\overline{\nb}$. Thus, analogically to definition of $\tau_{luu},$ the $\tau_{rnn}$ tensor is defined by

\begin{equation}
\label{lst2}
\tau_{rnn}=\sigma\left( \overline{\nb\nabla_s\cdot\nb\delta_S}-\overline{\nb}\nabla_s\cdot\overline{\nb}\right).
\end{equation}

Constant surface tension $\sigma$ is assumed. The $\tau_{rnn}$ is clearly a vector force, and will be nonzero only on the interface $S.$ Introduction of $\tau_{rnn}$ into (\ref{ff5}) results in 
\begin{eqnarray}
\label{ns+tau}
\frac{\partial \overline{\ub}}{\partial t} + \nabla \cdot \left(\overline{\ub}\otimes\overline{\ub}+\tau_{luu}\right) & = & \nonumber \\
\frac{1}{\rho}\nabla\cdot\left(-\overline{p}+\mu\overline{\db} +\sigma\overline{\nb}\nabla_s\cdot\overline{\nb}+\tau_{rnn}\right) +\mathbf{f}_g,
\end{eqnarray}
which serves as a LES-specific form of Navier -- Stokes equations applied for all calculations presented in this article. 

To summarize, we may say that two-phase LES is characterized by emergence of number of  tensors resulting from nonlinearity of filtered variables. First is the  $\tau_{luu}$ sub-grid stress tensor. Second is the term connected with phase-indicator transfer (SMT), and is created by filtering of the phase-indicator advection equation.  Finally, the sub-grid curvature tensor $\tau_{rnn}$ which has been formulated above.

\section{Solving Navier -- Stokes Equations}

Applied Navier -- Stokes solver was SAILOR-LES  \cite{sailor_jet, sailor_bif, sailor_elsner} , a projection-based high order code, utilizing a pseudo-spectral and 4th order compact discretisations for spatial, and low-storage Runge-Kutta schemes for temporal discretisation. LES approach has been applied using Smagorinsky sub-grid model for the $\tau_{luu}$ tensor in the oil-water mixing case described below\footnote{ The choice of this sub-grid model (the simplest of many implemented within the code) is motivated by the fact that the sub-grid stress tensor $\tau_{luu}$ was not the subject of our main interests in this paper, nor was it measured for comparison. Instead, the modelled $\tau_{rnn}$ tensor has been compared with resolved inertial terms, as will be discussed below.}. To advance the interface, the CLSVOF method has been applied, as described e.g. by Sussman  \cite{sussman1}  or Menard  \cite{menard}. Ghost-Fluid (GFM) technique  \cite{fedkiw-gfm} is used to couple the CLSVOF two-phase module with the solver. To facilitate calculations of two-phase flows, the SAILOR-LES solver includes the Multigrid  \cite{brandt} and the BiConjugate Gradients  \cite{fletcher} techniques to solve the Poisson equation.

\section{Advancing the Interface}

The CLSVOF  \cite{sussman1, tomar2005, gaurav} method is based upon simultaneous advection of the interface using VOF and LS methods, which allows for corrections of the Level-Set distance function using VOF distribution. It allows for substantial improvement in traced mass'  conservation over 'pure' LS methods  \cite{menard}, also  its accuracy concerning surface tension calculation it is comparable with modern VOF approaches  \cite{popinet2}  and has been applied to cases such as bubble growth  \cite{sussman2006,tomar2005} or jet atomisation  \cite{menard}. 

In this section, the description of VOF and LS methods is shortened to a minimum, while CLSVOF is described in a detailed way; this is because our implementations of VOF and LS methods are standard, while certain differences may be found between our implementation of CLSVOF and published works  \cite{sussman1,sussman2006}.

\subsection{Level Set Method}
The method is based upon the  Level-Set function $\phi$, which is equal in every point $\xb$ to the minimum distance $d$ between $\xb$ and the interface $S,$ so $\phi\colon=d(\xb,S).$ The function $\phi$ is therefore a \textit{distance function}. The function is advected using the fact that its material derivative must vanish:

\begin{equation}
\label{phiadv}
\frac{D\phi}{Dt}=\frac{\partial \phi(\xb,t)}{\partial t} + \ub(\xb,t) \nabla \cdot \phi(\xb,t)=0.
\end{equation}
It is assumed, that the traced interface is the zero-level of the $\phi$ function. Note that it means that we have an \textit{implicit} interface representation, as the actual localization of zero-level set is not necessary to solve the  advection equation. In Level-Set  methods  \cite{losasso, fedkiw3, fedkiw2,s-h}, this equation can be solved directly, provided that accurate discretisation is used (ENO and WENO  \cite{osherfedkiw} schemes have been proposed in this context and WENO has been implemented in our code). During the advection, due to numerical errors, $\phi$ will in general lose its distance property. Therefore it must be reinitialized, which requires solving the redistancing equation

\begin{equation}
\label{ls4}
\frac{\partial \phi}{\partial t'}+\mbox{sgn}(\phi_0)(|\nabla\phi|-1)=0,
\end{equation}
every few timesteps. In (\ref{ls4}), $t'$ is an internal re-initialization pseudo-timescale, and $\phi_0$ is the function distribution before the reinitialisation. 

Both advection and reinitialisation are known to cause changes in interface shape. Especially reinitialisation is known to cause smoothing of the $\phi$ distribution, which in computational practice may cause significant mass loss, especially when following small droplets or thin films  \cite{fedkiw3}.  

In our implementation, the curvature is calculated directly from the level set by means of second order central differencing.
\subsection{The Volume of Fluid Method}
The Volume of Fluid (VOF) method is one of the best established methods for two-phase flows  \cite{hn, szdirect, tsz}. The method is built upon a conception of a control volume (grid cell) containing the traced fluid, with the quantity of fluid being expressed as an integral of traced fluid's characteristic function

\begin{equation}
\label{vof2}
C_{i,j}=\frac{1}{h^2}\int\limits_V\chi(x,y)dxdy,\mbox{ }(x,y)\in(i,j),
\end{equation}
and called the \textit{fraction function}. In here, we consider a two-dimensional example. This fraction function is discontinuous, and therefore direct solving of its advection equation is not feasible, as it leads to diffusion of the interface shape. 

The discretized form of the advection equation for fraction function reads

\begin{equation}
\label{vof7}
h^2\frac{\partial C_{ij}(t)}{\partial t}+\int\limits_\Gamma \ub\cdot\nb\chi(\xb,t)d\Gamma=0.
\end{equation}
If we denote as $F_{i+1/2,j}$ the amount of $\chi$ leaving the cell during $\Delta t$ time step through the right wall\footnote{ Analogically $F_{i-1/2}$ is left wall flux, while $G_{i,j-1/2}$ and $G_{i,j+1/2}$ are respectively bottom and upper wall fluxes in $y.$} we can discretize (\ref{vof7}) in time arriving at

\begin{eqnarray}
\label{vof8}
C^{n+1}_{i,j} & = & C^n_{i,j}+\frac{\Delta t}{\Delta x}\left(F^n_{i-1/2,j}-F^n_{i+1/2,j}\right) \notag \\
              &   & + \frac{\Delta t}{\Delta y}\left(G^n_{i,j-1/2}-G^n_{i,j+1/2}\right),
\end{eqnarray}
where $\Delta x=\Delta y=h$  and $h$ is the grid spacing in uniform discretization. The fluxes formula above is derived from (\ref{vof7}) by using also continuity equation (\ref{cont}). Obviously, when (\ref{vof8}) is rewritten for 3D simulation, an additional term appears for the front and back wall fluxes. In modern implementation of VOF method, the $F$ quantities (fluxes) are found geometrically, by the so-called interface reconstruction  \cite{szdirect, youngs84, pilliod}. This reconstruction is demanding for the programmer, especially in 3D, however it has become a contemporary standard, since it guarantees perfect mass conservation. In our code, full implementation of VOF including PLIC (Piecewise Linear Interface Calculation) reconstruction of the interface has been performed. 


The outgoing flux $F_{i,j,k}$ is an intersection of volume $u_{i+1/2,j,k}\Delta t\Delta y\Delta z$ containing right-hand side wall of cell\footnote{Provided that $u_{i+1/2,j,k}$ is positive, see  \cite{aniszewski, tsz, sz2000} for details.}, and geometrically represented volume $C_{ijk}$ confined under the interface $\{ \nb_{i,j,k},\alpha_{i,j,k}\}.$   In actual implementation, calculation of the flux volume (or area, in two-dimensional case) requires considering all possible interface positions and matching appropriate formulae for volume/area. However, this is considerably simplified when one notices that in every case, the area confined between the interface \[\{ \nb_{i,j,k},\alpha_{i,j,k}\}\] and the origin of the local coordinate set is always an intersection of a properly defined tetrahedron and the cell cube  \cite{tsz}. The procedures utilizing this observation have been described by Scardovelli \& Zaleski  \cite{sz2000} for rectangular grids.  Note that for a given interface position $\{ \nb_{i,j,k},\alpha_{i,j,k}\},$ one is now able to establish a $1\colon 1$ functional dependence with the traced phase volume contained within the cell. So, when we fix the normal vector, we get 
\begin{equation}
\label{valpha}
V=V(\alpha)\end{equation}
functional dependence, whose properties are discussed in  \cite{sz2000, tsz, aniszewski} (this function can be reversed, to form $\alpha=\alpha(V)=\alpha(C_{ijk}),$ which is essential for interface reconstruction in modern PLIC approach). Also, we emphasize that finding flux volumes for equation (\ref{vof8}) is realized by using (\ref{valpha}) with transformations of local coordinates within the cell.

As it will be mentioned below, (\ref{valpha}) is employed within our CLSVOF implementation.


\subsection{Implementation of the CLSVOF algorithm}
     CLSVOF algorithm is relatively complicated, and is usually  \cite{sussman1} explained using numbered lists of algorithm steps. In here, we describe our implementation in a detailed way, together with a diagram of the basic method steps. In our approach, interaction between VOF and LS interface representations starts with generating an initial LS distribution, we then proceed along the stages described below. 

\subsubsection{Comparing Level Set and VoF Interfaces}
\begin{figure*}[ht!]
\centering
\includegraphics[scale=0.6]{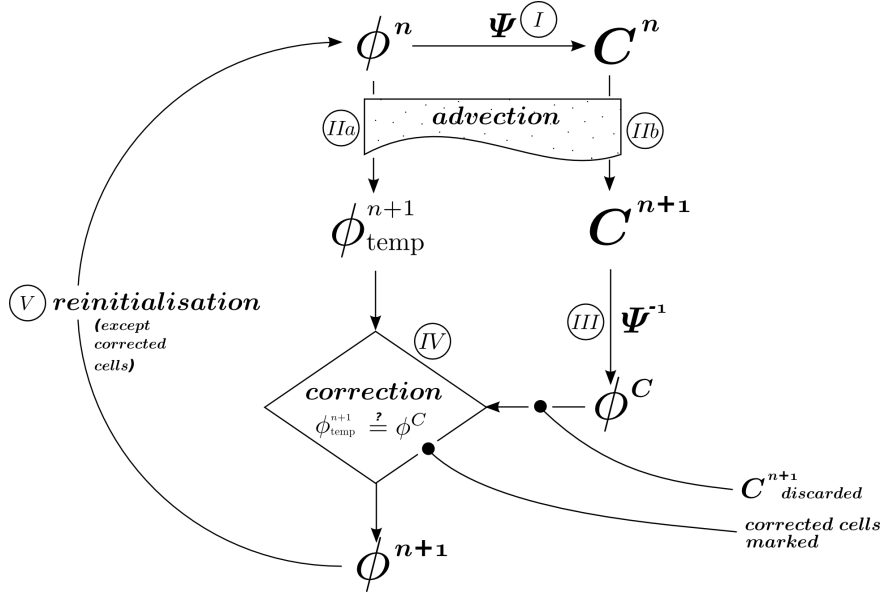}
\caption{Diagram of a CLSVoF algorithm.}\label{cls_diagram}
\end{figure*}

     As it is evidenced by diagram in Fig. \ref{cls_diagram} which should be read starting from upper left corner, the algorithm starts with $\phi^n$ being a level set from previous time step (obviously, $n=0$ at the code initialization). Then, as in 'pure' LS method, $\phi^n$ is advected with timestep $dt,$ thus procuring a $\phi^{n+1}_{temp}$ distribution (IIa on Fig. \ref{cls_diagram}). This is nothing else than $\phi^{n+1}$ one would obtain in pure LS method, yet in CLSVOF it is subject to further operations, hence the subscript. In order to enable juxtaposition of $\phi^{n+1}_{temp}$ with the effect of VOF advection, we first need to generate a (volume) fraction function $C^n$ from initial level set $\phi^n.$ This is done (step I) using a dedicated function.

Let us define a function 
\begin{equation}
\label{Psi}
 \Psi\colon \mathbb{R}^{\gamma}\mapsto\mathbb{R}^{\gamma}, 
\end{equation}
(where $\gamma$ stands for the dimension\footnote{Two dimensional $\gamma=2.$ situation is chosen for simplicity of description.}) such that

\begin{equation}
 \label{whatpsiis}
\Psi(\phi_{ij})=C_{ij}.
\end{equation}
Thus, $\Psi$ will be a function we need to make the comparison possible, as it has the following properties\footnote{For current examples, we will assume that level set function $\phi$ takes negative values $\phi < 0$ outside the traced phase, and is positive ($\phi >0$) inside; this choice is purely arbitrary. Besides, cell indexes are dropped.}:
\begin{itemize}
 \item $\phi < -\frac{\sqrt{2}}{2} \Rightarrow (\Psi(\phi)=C=0).$
 \item $\phi > \frac{\sqrt{2}}{2} \Rightarrow (\Psi(\phi)=C=1).$
 \item $|\phi| < \frac{\sqrt{2}}{2} \Rightarrow (0 < \Psi(\phi)=C <1).$
\end{itemize}

In the first case, considered computational cell is empty, in the second it is full, and in the third case it is a cell cut by the interface. To actually define $\Psi$ function we must specify how $C$ is computed in nontrivial cells. 


\textit{If} we can approximate implicit Level-Set interface with a straight line, and represent the line with $m_x x+m_y y=\alpha$ equation, we get the dependence

\begin{equation}
 \label{stephanessecret}
 \alpha = \frac{\phi}{m^{\phi}_x+m^{\phi}_y}+\frac{1}{2}
\end{equation}
where $m^{\phi}_{(\cdot)}$ denotes components of vector \textit{normal to the zero-isoline of the Level Set function}. As mentioned above, these normals, given $\phi$ distribution, can be easily found e.g. using Youngs' scheme.  This way, using (\ref{stephanessecret}) in nontrivial cell, one can find values of $\alpha$ in a nontrivial cell. \footnote{The division by level set normals is motivated by condition $m_x+m+y=1$ held for positive $m_x$ and $m_y$, under which $\alpha$ can be mapped onto $\lbrack 0,1 \rbrack.$}

We are now able to find $\alpha$ and $\nb^\phi$ for LS interface. We can now use the flux-calculating procedure (dependence (\ref{valpha})), with $u\delta t=1$ to find the quantity of traced mass delimited by the  interface.  This will yield the total area/volume under the interface, and $\Psi$ will be defined. So to reiterate, $\Psi$ is a combined function, which written with all arguments list would have the form:

\begin{equation}
\label{complete_psi}
\Psi(\phi)=Flux(\Psi((\nb^\phi(\phi),\alpha(\nb^\phi(\phi))),\mathbf{u},\delta t))
\end{equation}
using $u\delta t=1$ and with $Flux$ being a VOF-specific function (\ref{valpha}) used to calculate the fluxes ( in (\ref{vof8})) of $C,$ through the walls. This approach seems simpler than least-square minimization approach of \cite{menard}.


Being equipped with $\Psi$ function, we can generate $C^n$ from $\phi^n$ and by this finish step I  (refer to Fig. \ref{cls_diagram}).  Step II.a is an aforementioned advection of $\phi.$ Now II.b is a VOF advection carried out in a classical manner, just as it would be done in normal VOF-PLIC implementation. In effect, we get $C^{n+1},$ an advected VOF interface. 

The point of CLSVOF -- at least in its original formulation -- is to compare LS and VOF distributions  to enable correction. However $\phi$ and $C$ are only comparable through $\Psi(\phi)=C,$ hence we need $\Psi^{-1}.$ This is step III, using  
\begin{equation}
\label{correct1}
\phi^C=\Psi^{-1}|_\Gamma(C^{n+1})
\end{equation} 
to generate a level set function directly comparable with $\phi^{n+1}_{temp}.$ The $|_\Gamma$ designates restriction to the interface. The restricted reverse of $\Psi$ is in fact designed much like the original function, due to the bijection between $\phi$ and $\alpha(C,\nb)$ that exists on $\Gamma.$ Therefore $\Psi^{-1}\colon \Gamma\rightarrow \lbrack -d,d\rbrack$ and
\begin{equation}
\label{correct2}
\phi^C=\Psi^{-1}|_\Gamma(C^{n+1})=\alpha(C^{n+1},\nb(C^{n+1}))-\frac{1}{2},
\end{equation} 
and $d$ is the diagonal of a computational cell.

 Construction of $\Psi^{-1}$ on a broader domain is highly nontrivial (although valuable attempts exist i.e by Cummins et al.\cite{cummins}). Above function allows to calculate $\phi^C\colon=\phi(C^{n+1})$ in interface cells, and for such cells, we will have now clear view of possible differences $\phi^C-\phi.$

The next stage is step IV, i.e. the correction of the level set.

\subsubsection{Correction of the Level Set}

To correct the values of Level Set, one needs correction criteria, allowing to choose (mark) cells that require correction.  We now have computed $\phi^C, \phi^{n+1}_{temp}$ and $C^{n+1}.$ Also, at this point, which on Fig. \ref{cls_diagram} is marked IV, we generate $C^{\phi}=\Psi\left(\phi^{n+1}_{temp}\right).$  In our current implementation, the values of $\phi$ are corrected under the following conditions:
 \begin{enumerate}
  \item $|\phi| < d/2=h\sqrt{2}/2$
  \item $0 + \epsilon < C^{n+1} < 1-\epsilon$
  \item $\phi^{n+1}_{temp}\ne \phi^C$, or $C^\phi \ne C^{n+1}$
 \end{enumerate}
with $C^{\phi}$ described above.   First requirement ensures, that $\phi$ changes sign inside the cell - as $ d/2 =h\frac{\sqrt{2}}{2}$ is the maximal distance from (square) cell of size $h$ to its corner. In other words, ensures that interface is passing through the cell. In general, for cell of size $\Delta x \times \Delta y \times \Delta z,$ we need half of the cuboid's diagonal $\frac{d}{2}=\sqrt{\Delta x^2+\Delta y^2+\Delta z^2.}$

Second criterion, with $\epsilon$ being a small number, is a simple requirement of VoF fraction function non-triviality.  This way, even if the limit in criterion 1 is too strong, we can evade treating trivial (full/empty)  cells that probably don't need correction. If $1-C^{n+1}<\epsilon$ (such cell is considered full) and at the same time $\phi$ indicates empty cell, the correction will not take place. This situation is however considered unlikely as it assumes that a very large discrepancy between VOF and LS interfaces can be created during one timestep. 

Third criterion is based on our previous discussion, that values of $C$ originating in VoF advection are more reliable that $C^{\phi}.$ Actually, we can allow these two values to be reasonably close, and correct Level Set only if the difference is bigger than a set value. Authors of  \cite{sussman1} advice using $|C^{\phi}-C| < 0.001$ as a criterion.  Also, some of the criteria may not be used at all (or loosened), increasing the number of corrected cells, though it's not profitable to correct in \textit{every} interface cells because it degrades the quality of curvature calculated from level set.

As we have found in our experiments, setting the right criteria is a matter of certain delicacy - ideally only cells in under-resolved areas should be corrected. Setting too broad criteria will cause $\phi$ values to change everywhere on the interface, while very strict levels may cause no correction at all. 

After deciding that a certain cell will be corrected (Fig.\ref{cls_diagram}, step IV), we mark it and then simply change its $\phi$ value to $\phi_c.$ This finishes most of the procedure; the only remaining step is the reinitialisation/redistancing of $\phi$ (step V). The important part of CLSVoF algorithm is the omission of marked cells (the ones that underwent correction), so that their $\phi$ values will not be changed during the reinit. 

It is important to note, that in our implementation of the CLSVoF algorithm, the $C$ function is \textit{not} conserved throughout the simulation, but rather recreated from $\phi$ at the beginning of each timestep. This way, the traced mass may  still be lost in reinitialisation of $\phi,$ we adress this problem by modifying the redistancing procedure\cite{sussman1, menard}, however the mass consvation is still inferior to ``pure'' VOF method. The advantage of CLSVOF is however the possibility to utilize the Ghost Fluid technique.

 \section{Approximate Deconvolution}\label{admsect}

The Approximate Deconvolution Model (ADM) was developed by Stolz et al.  \cite{adams} and applied to incompressible wall-bounded flow. The idea of ADM is to recreate the filtered sub-grid scales of the flow. If the filtering operation is denoted by convolution kernel $G,$ we could denote (for a compact support filter):

\begin{equation}
\overline{u(x)}=G\star u=\frac{1}{\Delta}\int\limits^{x+\Delta/2}_{x-\Delta/2}G(\frac{x-x'}{\Delta},x) u(x')dx',
\label{adm1}
\end{equation}
where $\Delta$ is the filter width. In shortened notation using new variable $z=\frac{x-x'}{\Delta}$ we get

\begin{equation}
\overline{u(x)}=\int\limits^{x+\Delta/2}_{x-\Delta/2}G(z,x)u(x-z)dz.
\label{adm2}
\end{equation}
The three-dimensional form is obtained by filtering independently along three coordinate axes, i.e. as a convolution of three filters $G_x, G_y$ and $G_z$ defined analogously to (\ref{adm1}). Furthermore, commutation is assumed between filtering and derivation operations. 

The idea of ADM is to calculate $\tau_{l u u} $ by a simple model - directly , using $u_i^*,$ an approximation of unfiltered quantity $u_i.$ It is therefore a deconvolution class closure, similar to Domaradzki's approach  \cite{domaradzki}. The deconvolution technique originated in computer graphics, where it was using for 'sharpening' images. One may think of reconstructed function $f$ as being a ``sharpened'' $\overline{f}.$ 

Using $u_i^*$ the modeled tensor reads simply:

\begin{equation}\label{adm6} \tau_{lu u} = \frac{\partial\overline{ u_j^* u_i^* }}{\partial x_i}-\frac{\partial \overline{u_i}\hspace{2pt}\overline{u_j}}{\partial x_j}.\end{equation}
Using the above formulas, ADM proceeds to replace $u_i$ with $u_i^*$ in nonlinear terms. It has to be noted that to ensure proper energy drain from resolved scales, original ADM proposed by Stolz et al.  \cite{adams} uses also the relaxation term

\begin{equation}
-\chi(I-G^{-1}\star G)\star \overline{u_i}
\label{adm9}
\end{equation}
added to the right-hand side of (\ref{ns+tau}), with $\chi_u$ being relaxation parameter and $G^{-1}$ denotes the inverse of filter $G.$ Since, as will be explained below, $G^{-1}$ is approximated, this term generally is nonzero. Calculation of (\ref{adm9}) has been therefore implemented in the code used for present study.  

The approximate inverse of G is based upon analogy between functionals over $\mathbb{L}^2$ space and real functions over $\mathbb{R},$ namely the analog of Taylor expansion. Existence of such a converse of Taylor Theorem has been investigated for example by Dayal \& Jain  \cite{dayal}.

Approximate inverse $G^{-1}_N$ (of $N$th order) of $G$ is, assuming its existence, expressed by

\begin{equation}
G^{-1}_N=\sum\limits_{l=1}^N (I-G)^l,
\label{adm10}
\end{equation}
provided that $||I-G||<1,$ i.e. the series is convergent.
In practical application, the expression for $G^{-1}$ is truncated. Adams suggests that setting $N=5$ gives acceptable results. For example, expanded expression for $N=4$ would be

\begin{equation}
G^{-1}_4=4-10G+10G^2-5G^3+G^4,
\label{adm11}
\end{equation}
so that

\begin{equation}
u^*=G^{-1}_4\star \overline{u}=4\overline{u}-10G\star \overline{u}+10G^2\star \overline{u}-5G^3 \star \overline{u}+G^4\star \overline{u},
\label{adm12}
\end{equation}
where $G^2\star u =G\star(G\star u).$ 

Choice of a particular filter for ADM could, as suggested by Geurts  \cite{geurts}, be linked to the fact that spatial discretisation leads to behaviour reminiscent of filtering. It means, that any spatial discretisation induces filtering of all quantities (derivatives) calculated by with it. In that manner Geurts shows an example of first order finite difference scheme inducing a top hat filter of width equal to grid spacing.  If that were so while one uses grid-based LES, it would be desirable that filters used for ADM mimicked the filter-inducing behaviour of spatial discretisation. Intuitively, it would mean that ADM reverses the filtering induced by discretisation. However, filter construction technique proposed in   \cite{geurts} applies only to explicit discretisations, such as finite differences, while SAILOR-LES uses a P\'{a}de and pseudo-spectral schemes, with an option to mix the two.  We therefore are yet unable to propose analytic formula for discretisation-induced filter $G$ relevant to our simulation, instead we use the explicit fourth order filter that applies a five point stencil, proposed by Stolz et al.   \cite{adams,jeanmart}

\begin{eqnarray}
\overline{f(x)} &= &f(x)-(f(x+2h)-4f(x+h)+6f(x)\nonumber \\ 
                &  & -4f(x-h) +f(x-2h))/16
\label{adm15}
\end{eqnarray}

\section{Calculation of $\tau_{rnn}$ Tensor}\label{tau_section}

The capillary tensor, resulting from unresolved interface shape is expressed  with (\ref{lst2}). The $\delta_S$ is computed using condition on the level set, such as 
\begin{equation}
\label{tau_delta}
\delta_S(\mathbf{x})=1 \Leftrightarrow |\phi(\mathbf{x})|<\min(\Delta x,\Delta y,\Delta z).
\end{equation}
The $\nabla \cdot \ndash$ term in (\ref{lst2}) is curvature of the interface  calculated from the level set. 

As with $\tau_{luu},$ modelling of $\tau_{rnn}$ tensor is required since $\mathbf{n}$ is unknown in a Large Eddy Simulation. Thanks to the ADM algorithm, it is however possible to produce quantity that we define as $\mathbf{n}^*,$ a~vector field of reconstructed normals. This can be achieved in (at least) two different approaches tested during the preparation of this article, which we will now describe:

\begin{itemize}
\item[(A)] Direct ('explicit') deconvolution of $\mathbf{n}^*,$ i.e. $\mathbf{n}^*=G^{-1}(\overline{\mathbf{n}})$ where $G^{-1}$ is the deconvolution operator.
\item[(B)] Indirect ('implicit') deconvolution, based upon assumption that for a given velocity field $u_i(x_i,t),$ the $\mathbf{n}$ field can be represented as dependent from the velocity field
  \begin{equation}
    \label{tau2}
    \mathbf{n}=f(u_i(x_i,t)),
  \end{equation}
  and $f$ is an injective function\footnote{That is, if $a,b\in X$ then $a\neq b \Rightarrow f(a)\neq f(b).$} . In other words there is at least injective dependence between velocity field and the interface shape\footnote{Expecting $f$ to be a bijection would be, intuitively, logical with zero boundary conditions and no heat exchange, i.e. in case of flow driven by surface tension and density jump.}, which seems justified, since the only way for the interface to change its shape is to undergo advection\footnote{It is, however, implied, that $\mathbf{n}^*=\overline{\mathbf{n}}$ whenever there is no flow, e.g. in case of stationary spherical droplet. }. This way $\nstar$ calculation is realized by finding $\ustar.$
\end{itemize}

Since in (B) approach $\ustar$ is the approximately deconvoluted $\ub,$ we could follow original approach of Stolz et al.  \cite{adams} and include $\ustar$ in simulation effectively discarding any sub-grid models used in SAILOR. In this work, however, $\ustar$ is used only for interface advection. 

 The (B) approach was chosen and applied for most of calculations. The rationale for this is that the deconvolution of the velocity field seems less prone to errors resulting from oscillations caused by high-order filters when processing variables with jumps, such as would inevitably appear in case of normal vectors, which are defined only on the interface. This could subsequently cause errors in resulting $\tau_{rnn}$ values. 

The (B) variant is implemented as follows. Advection of the interface is carried out twice, once using velocity field $\udash,$ yielding $\ndash$ and once again with $\ustar$ resulting in $\nstar.$ Resulting quantities are stored in separate tables and used later to calculate $\tau_{rnn}$  using its definition (\ref{lst2}). 

In the CLSVOF method, there exists a possibility to obtain at any moment both normal vectors calculated from $C$ and from $\phi$ function. Therefore, similar to CSF approach of Brackbill et al. \cite{brackbill}, VOF  could be used to obtain $\nabla\cdot \nb$ in (\ref{lst2}). In our calculations however, the curvature calculation in the Ghost-Fluid procedure utilises $\nb(\phi),$ as the LS method guarantees smooth representation of the interface \cite{fedkiw-gfm}.  Therefore, for consistency, we are also using level set-derived normals to calculate $\tau_{rnn}$ in (\ref{lst2}).

\begin{figure}[ht!]
\centering
\includegraphics[scale=0.5]{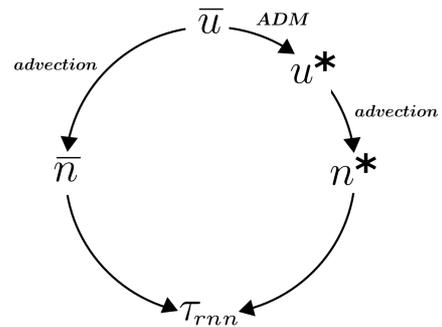}
\caption{Diagram of the $\tau_{rnn}$ calculation algorithm, (B) approach.}\label{tau_diagram}
\end{figure}

  Simple diagram of the $\tau_{rnn}$ reconstruction  procedure can be seen in Fig. \ref{tau_diagram}. Our computational practice shows that using high-order solver with Conjugate Gradients and Multigrid techniques such as SAILOR, the computational cost of running CLSVOF twice is still merely a fraction of CPU time needed for Poisson equation.

\subsection*{Remark Concerning the ``Sub-Grid Mass Transfer''}

Toutant  \cite{toutant} and Vincent  \cite{vincent} describe, besides $\tau_{rnn}$ tensor mentioned here, another term -- which is entirely of numerical origin -- and emerges when VoF-specific $C$ function transport equation is subject to filtering, namely 

\begin{equation}
\label{tau3}
\frac{\partial \overline{C}}{\partial t} + \udash\cdot\nabla\overline{C}=0,
\end{equation}
contains nonlinear convective term, to which attributed is the term 
\begin{equation}
\label{tau4}
\sigma_1=\ub\cdot\nabla\overline{C}-\overline{\ub\cdot\nabla C}
\end{equation}
which we denoted $\sigma_1$ following  \cite{vincent} and refer to it as a \textit{sub-grid mass transfer} (SMT) term, since it involves unfiltered (sub-grid) distribution $C.$  Obviously, usage of levelset $\phi$ function instead of $C$ does not change anything in this context.

In present authors' opinion, a similar deconvolution-based approach could be feasible for SMT term, however in the  present work, modelling of $\sigma_1$ has not yet been performed. We agree that it should be subject to future study, as  \cite{vincent} proves that the term has high values in computational case of phase-inversion problem (Section \ref{results_mixing_section}), relatively to resolved convective term in (\ref{tau3}).

\section{Numerical Experiments}

\subsection{Advection Scheme Testing}\label{results_coupling_section}

Some results of three dimensional CLSVOF coupling performance (without ADM model for $\tau_{rnn}$) are presented here.
To begin, let us consider a velocity field $\ub=(u,v,w)$ defined by:

\begin{eqnarray}
\label{bfield}
u(x,y,z,t) & = & 2\sin^2\pi x \sin \pi y\sin \pi z\cos\pi t \nonumber \\
v(x,y,z,t) & = & -\sin \pi x \sin^2 \pi y \sin \pi z\cos\pi t \nonumber \\
w(x,y,z,t) & = & -\sin \pi x \sin \pi y \sin^2 \pi z\cos\pi t.
\end{eqnarray}  

This field  \cite{menard}  presents eight artificial vortices in domain octants. Let $x,y,z\in \lb 0,1 \rb,$  and $t\in (0,1),$ actual role of $\cos\pi t$ coefficient is to change sign when $t=\frac{1}{2}.$ Spherical droplet of radius $r=\frac{1}{5}$ is placed inside the domain at point $(\frac{1}{3},\frac{1}{3},\frac{1}{3}).$ Under such conditions, passive advection of the droplet using (\ref{bfield}) is performed. This test is widely presented in subject literature  \cite{menard, fedkiw3,berlemont}, and its purpose is to introduce substantial droplet deformation, by which method's ability to trace thin films, topology changes and its mass conservation capabilities are assessed. If $t$ is allowed to change in aforementioned manner, velocity field will reverse and the droplet will go back to it initial position. However, due to numerical errors, every method introduced some form of an error. Comparison of initial and final droplet shape is therefore -- once again -- a possibility to qualitatively and quantitatively check method errors. 

\begin{figure}[ht!]
\centering
\includegraphics[scale=0.18]{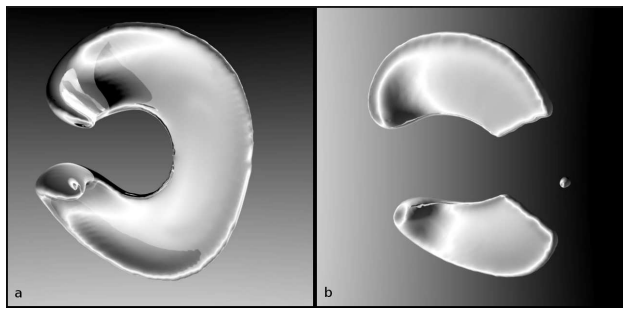}
\caption{Comparison of method performance: CLSVOF advection (a) and LS advection (b) of a circular droplet under velocity field (\ref{bfield}).}\label{berl_comp}
\end{figure}

Let us also notice that for (\ref{bfield}), divergence of $\ub$ is analytically nonzero, causing the assumption of $\ub$ being solenoidal is unfulfilled in derivation of VoF (equation (\ref{vof8})). This in turn may introduce $C$ values outside the interval $\lb 0,1 \rb.$ Combined with ``clamping'' procedure -- in which values of $C$ are restricted to $\lb 0,1 \rb,$ -- which is commonly used in split-advection VOF implementations, this may lead to loss of traced mass. In practice, approximately $5\%$ of mass has been observed to be lost for advections carried to the stage depicted in Fig. \ref{berl_comp} on coarse $32^3$ grid, with smaller values on finer grids. In general, to prevent this, the correction for the divergence is addded to (\ref{vof7}), e.g. in  \cite{xiao}. 

When $t$ is disregarded, field (\ref{bfield}) is strictly steady  \cite{batchelor}. Therefore, it has static streamlines, that is solutions of \begin{equation} \label{bstreamlines} \frac{dx}{u(\xb)}=\frac{dy}{v(\xb)}=\frac{dz}{w(\xb)}.\end{equation} 

\begin{figure}[ht!]
\centering
\includegraphics[scale=0.27]{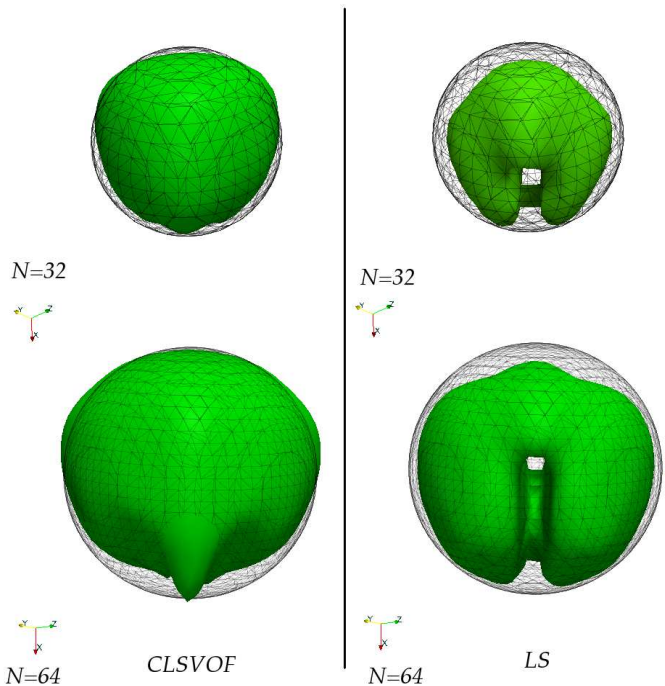}
\caption{Final shape of the droplet, for CLSVOF and LS methods. Wireframe delineates original ($t=0$) shape, while green isosurface depicts final shape ($t=\pi$)}\label{berl_back}
\end{figure}

Figure \ref{berl_comp}(a) presents the shapes of distorted droplet, after it has been advected under (\ref{bfield}) and in (b) the  result of similar simulation using a pure Levelset method. Droplet shape is presented for $t=\frac{\pi}{2},$ that is exactly at the moment when droplet is most deformed. A thin film of fluid is traced by CLSVOF\footnote{Analysis of layer thickness is widely covered in  \cite{menard}.}, in contrast to LS, where respective part of mass has been lost. In addition, Figure \ref{berl_back} presents the final shapes of the droplets obtained at $t=\pi$ for the very coarse ($32^3$) and medium-sized ($64^3$) grids, enabling the qualitative comparison of results (in Fig. \ref{berl_back}, left-hand-side pictures correspond to the results presented in second and third line of  Table \ref{berlemont_table}).


For $t=\pi,$ the droplet should have returned to its original shape. 

\begin{table}
\begin{center}
\caption{The $L_1$ error for 3D CLSVOF passive advection using velocity field (\ref{bfield}).}
\begin{tabular}{|c|l|c|c|}
\hline
$N$ & $L_1$ error & rate & order \\
\hline
16 & 0.03371 & -- & -- \\
32 & 0.0197 &  1.71  & 0.855\\
64 & 0.00838 & 2.35 & 1.175 \\ 
128 &  0.00326  &  2.56  & 1.28\\
\hline
\end{tabular}
\label{berlemont_table}
\end{center}
\end{table}

To further investigate the performance of the CLSVOF method in passive advection test, we perform the following test using velocity field (\ref{bfield}). The droplet was advected from, and returned to its starting position, since the term $\cos\pi t$ in (\ref{bfield}) changes sign as $t>0.5.$ Initial and final droplet shapes differ, depending on the performance of the advection method applied. The difference may be quantified using $L_1$ error.   To present exact error formulation, let us introduce following symbols: $N$ as a number of grid nodes in uniform grid (with $N^3$ nodes), $(i,j,k)$ as index of grid cell, with all three indices ranging from $1$ to $N.$ Then

\begin{equation}
\label{l1berl}
L_1=\frac{1}{N^3}\left(\sum\limits_{i,j,l=1}^N |C^0_{ijk}-C^f_{ijk}|\right),
\end{equation}
with $C^0$ standing for the initial, and $C^f$ the final distributions of the VOF fraction function $C.$ Normalizing by $N^3$ is a simple technique to assure that errors are comparable. Results of this numerical test are presented in Table \ref{berlemont_table}. The ``rate'' column is an estimation of error decrease between consecutive table rows  \cite{pilliod}, so that for example \[ \frac{L_1|_{N=32}}{L_1|_{N=64}}=\frac{0.0197}{0.00838}\approx 2.3\] is visible in third row. From this estimation, we can see that the method (that is, the advection scheme) is approximately first order in accuracy, $O(N).$ 

For the results presented in this subsection, Parker \& Youngs' \cite{youngs92} method is used for calculation of the normal $\nb$ to the interface. Therefore, first-order accuracy is expected for VOF method and fully consistent with the results of Pillod and Puckett \cite{pilliod}. 

\begin{figure*}[ht!]
\centering
\includegraphics[scale=0.85]{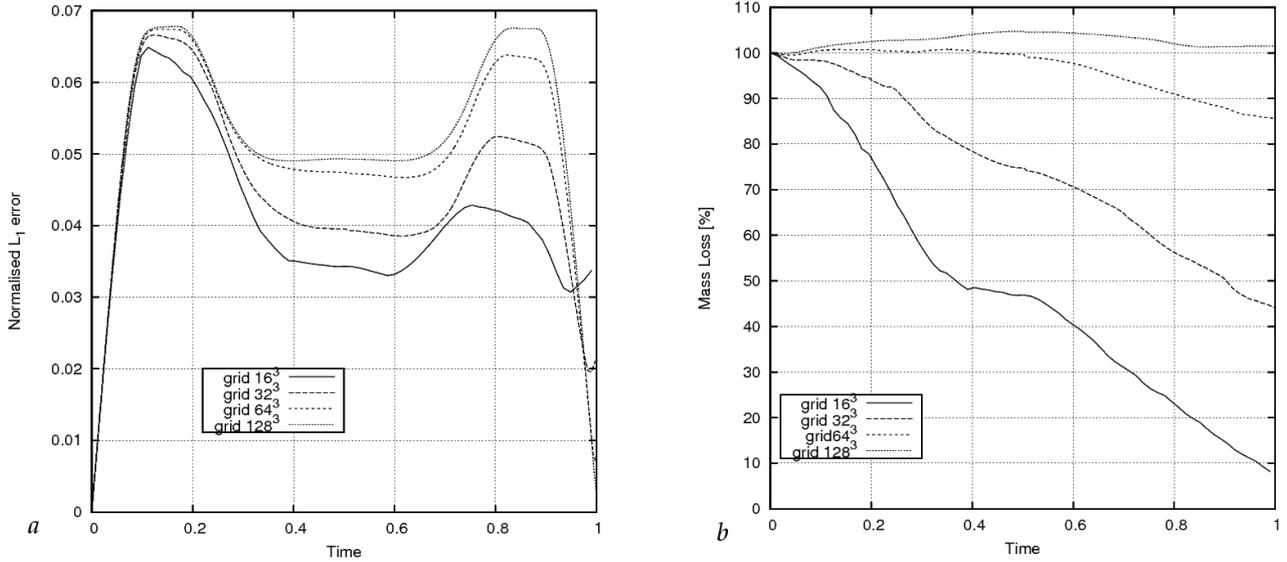}
\caption{ Temporal evolution of (a) errors, (b) normalized dimensionless mass for velocity field (\ref{bfield}).}\label{berr}
\end{figure*}

When plotted against the dimensionless time, the $L_1$ error (Fig. \ref{berr}a)  shows that for $N=16$ the curve does not return to zero value. This is caused by the fact that with this drastically coarse grid resolution, most of the mass have been lost by the time the droplet returns to its original position\footnote{The CFL condition is utilized to calculate timestep for each grid size. Hence, different timestep $dt$ is obtained and the number of numerical timesteps is not identical.}. Two peaks on Fig. \ref{berr}a are caused by geometry of field (\ref{bfield}), that is: the droplet leaves its initial location (first peak - high error), then it undergoes severe deformation, causing parts of it to coincide with its original location (error drops), which takes place for $t=0.5.$ After this, the process returns to its original state (second peak, and return of error to zero value). 

The issue of mass conservation in this case can be studied by inspecting Fig. \ref{berr}b, in which mass conservation is plotted, as the percentage of droplet's initial mass. Calculation of plotted value is made possible by calculating
\begin{equation}
\label{berlmass}
M(n)=\sum\limits_{i,j,l=1}^N C^n_{ijk},
\end{equation}
where $n$ stands for the $n$-th step of temporal discretisation. Value plotted in Fig. \ref{berr}b is therefore $M(t)/M(0).$
It is visible that while $N=64^3$ grid results in reasonable mass conservation ($90\%$ of initial mass), on the finest grids overshoots are created. Such a mass loss would seem unlikely in pure VOF method (where the sum (\ref{berlmass}) is constant by definition), but in CLSVOF it can be attributed to the process of re-creation of VOF distribution after every advection step. 

While values for mass conservation over $90\%$ may be perceived low compared to machine-precision VOF conservation, it exceeds values obtained by using LS method  \cite{osherfedkiw}. Moreover, results presented by other authors  \cite{menard,svof,fedkiw3,xiao} show that velocity field (\ref{bfield}) is very demanding -- when the ability of a given method to conserve mass is considered --  grids used by these authors are relatively fine, up to the level of $N=200.$

An additional, similar test has been performed using a simpler velocity field, namely
\begin{equation}
\label{sfield}
u(t)=v(t)=w(t)=\cos(\pi t),
\end{equation}
which was spatially constant. The droplet was placed in point $(0.3,0.3,0.3)$ of an $\lb 0,1\rb^3$ domain. Passive advection in (\ref{sfield}) is equivalent to movement of the droplet to the opposite corner of the domain, and then back to the original location. This simple example is an excellent test for VOF-based advection schemes, similarly to many tests basing on advecting droplets along axes, advection over periodic walls for many periods and similar cases  \cite{afkhami_phd}.

 Analogically to the previous tests, advection on four similar grids have been performed -- however, quantitive visualizations of these will be omitted, since the difference between initial and final droplet positions are very small. Instead, we present error analysis in Table \ref{shift_table}. From the table, it is visible that in this case a slightly higher error decrease rate is observed, nearing second order between grids $32^3$, however does not hold between grids $64^3$ and $128^3.$ This behaviour is comparable to  observed by Sussman et al.\cite{sussman1}, who describe CLSVOF algorithm that exhibits interchangeably first- and second-order accuracy in parts of temporal evolution of simulation.

\begin{figure}[ht!]
\centering
\includegraphics[scale=0.8]{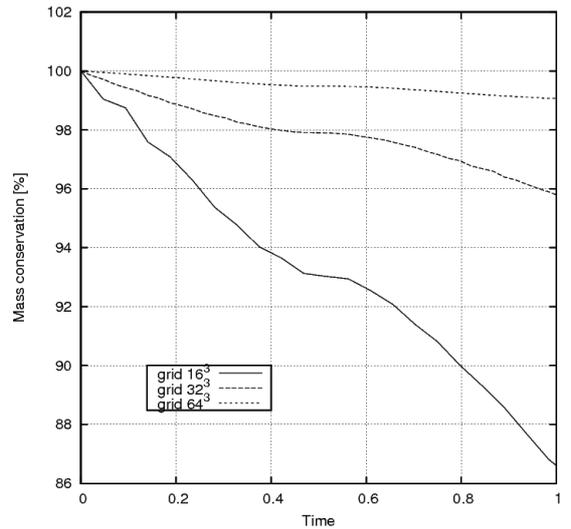}
\caption{Dimensionless mass  temporal evolution for velocity field (\ref{sfield}), normalized by the initial value.}\label{scurves}
\end{figure}

Unlike (\ref{bfield}), field (\ref{sfield}) has divergence zero, hence much lower mass loss is expected. Indeed, as can be observed in Fig. \ref{scurves} mass conservation for $64^3$ grid reaches $99\%,$ while even for the coarsest $16^3$ grid over $80\%$ of mass is conserved unlike in the previous test, when this very coarse grid performed on the $20\%$ level. Result for the $128^3$ grid is not pictured, since the value is constant and equal to $100\%.$

\begin{table}
\begin{center}
\caption{The $L_1$ error for 3D CLSVOF passive advection using velocity field (\ref{sfield}).}
\begin{tabular}{|c|l|c|c|}
\hline
$N$ & $L_1$ error & rate & order \\
\hline
16  & $4.625\cdot 10^{-3}$ & -- & --\\
32  & $1.442\cdot 10^{-3}$ & 3.207  & 1.6 \\
64  & $3.729\cdot 10^{-4}$ & 3.866  & 1.93 \\
128 & $2.511\cdot 10^{-4}$ & 1.485  & 0.724 \\
\hline
\end{tabular}
\label{shift_table}
\end{center}
\end{table}

\subsubsection{The Calculation of Curvature}

As was said, the curvature $\kappa$ is calculated directly from the level-set $\phi$ function by using a second order differences scheme. A geometrical analysis of the curvature calculation scheme was performed in two parts: a static test and ''parasitic currents'' analysis. For a static test, the curvature of a spherical $\phi$ distribution defined by setting $\phi=\sqrt{x^2+y^2+z^2}-0.1$ was found using our scheme, with the error defined from the knowledge of analytical curvature \begin{equation}\label{kappa_an}\kappa=\frac{1}{\sqrt{x^2+y^2+z^2}}.\end{equation} Such $\phi$ distribution corresponds to a sphere of radius $0.1$ however, it is obviously possible to calculate $\kappa$ (as well as calculate $\phi$ normal vectors) regardless of the position of the zero level-set. The sphere was placed in the centre of box domain of dimensionless size $L=1,$ using varying number of grid cells in each direction. Curvature measurement was performed at three radii chosen from $\lb 0, 0.3L\rb$ interval. The results, together with an order estimation, are presented in Table \ref{curvtable}.

\begin{table}
\begin{center}
\caption{Error calculation for static curvature test with order estimation.}
\begin{tabular}{|c|c|c|c|c|}
\hline
$N$ & 0.05L & 0.15L & 0.3L  & order \\
\hline
20  & $3.4 \times 10^0$   & $0.2 \times 10^0$ & $2.3 \times 10^{-2}$ & --\\
40  & $1.1 \times 10^0$   & $4.6 \times 10^{-2}$ & $5.8 \times 10^{-3}$ & 1.6\\
80  & $0.3 \times 10^0$   & $1.2 \times 10^{-2}$ & $1.5 \times 10^{-3}$ & 1.9\\
160 & $7.8 \times 10^{-2}$ & $2.9 \times 10^{-3}$ & $3.6 \times 10^{-4}$ & 1.95\\
\hline
\end{tabular}
\label{curvtable}
\end{center}
\end{table}

The investigation of ''parasitic currents'', which are the numerically induced non-zero velocities that arise due to errors of curvature calculation schemes \cite{popinet2} was performed in a following manner. A fully coupled SAILOR-CLSVOF code was used, with 3D domain of nondimensional size $L=5.0,$ using $64^3$ uniform grid. A spherical droplet of radius 1 was placed in the center of domain, defined by the initial $\phi$ distribution analogous to a 2D example presented above for curvature calculation. There was no jump in density ($\rho_1=\rho_2=1.$) nor viscosity ($\mu_1=\mu_2=0.005)$. Surface tension coefficient has been set to $\sigma=0.375,$ and gravity is zero. Under such conditions, we have accepted $max(|\ub|)$ measured in domain as an estimation of parasitic currents phenomenon. The maximum was found in whole domain at given moment of dimensionless time $t.$

\begin{figure}[ht!]
\centering
\includegraphics[scale=0.17]{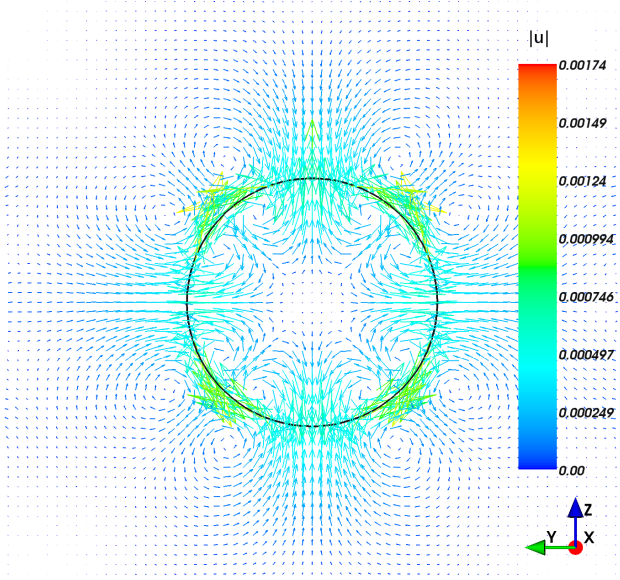}
\caption{Vector cutplane of the parasitic current distribution for $t=0.211\cdot10^2.$}\label{par_vf}
\end{figure}

Symmetrical distribution of the parasitic currents obtained for the CLSVOF code at $t=0.211\cdot10^2$ can be seen in Figure \ref{par_vf}. As can be observed, most of the visible vectors are of order $10^{-4},$ while the domain-averaged $|\ub|$ value was of order $10^{-7}.$ This is comparable with published results \cite{menard} and \cite{spc}\footnote{Where similar Ohnesorge number was analised, but 2D advection was considered.}, also relates well to data available in \cite{tsz}. 

\begin{figure}[ht!]
\centering
\includegraphics[scale=1.42]{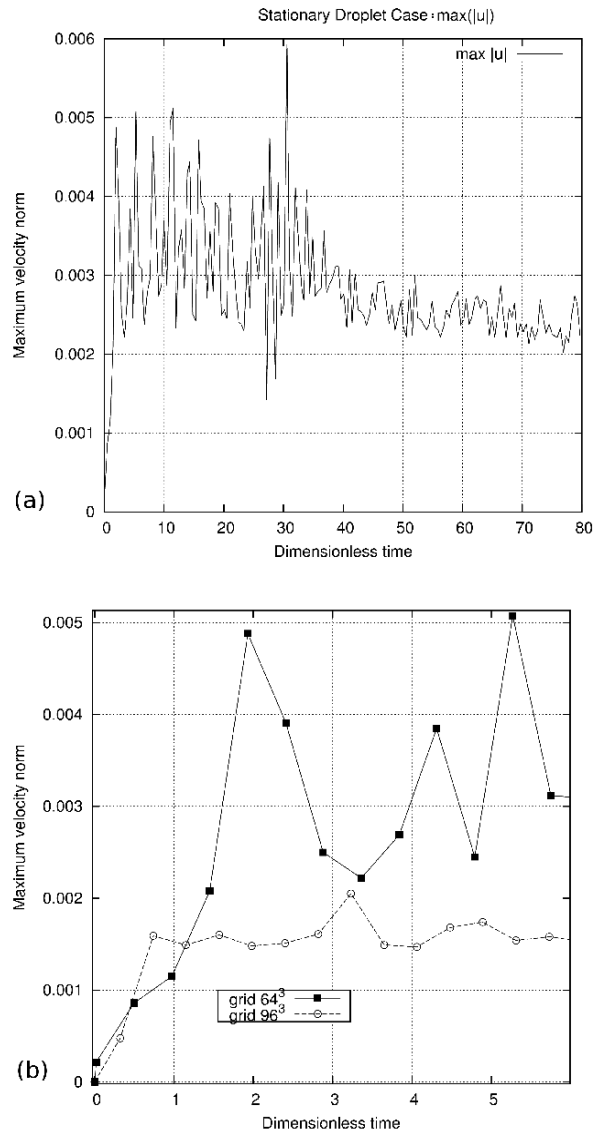}
\caption{Temporal evolution of spurious velocity maximum for the stationary droplet test case using $64^4$ grid.}\label{par_te}
\end{figure}

Temporal evolution of the $max(|\ub|)$ can be observed in Fig. \ref{par_te}a, prepared on a $64^3$ grid. As can be seen in this Figure, oscillations occure which are dumped in the presence of viscosity. Average value of $max(|\ub|)$ is of order $10^-3$ with a decreasing tendency as $t$ progresses. The plot was prepared using aproximately $5*10^3$ iterations of the solver, with each value for $max(|\ub|)$ searched in entire domain and saved every 10 iterations. This maximum values are consistent with data presented in Figure \ref{par_vf}. Additionally, Figure \ref{par_te}b presents a detailed view of temporal $max(|\ub|)$ evolution for $t\in\lb 0,10\rb,$ for both $64^3$ and  $96^3$ grids. A substantial decrease by approximately $60\%$ in  median value of $max(|\ub|)$ is seen as  the grid changes from $64^3$ to $96^3.$

\subsection{The Phase Mixing}\label{results_mixing_section}
Much attention was devoted by the present authors to the computational case of  phase mixing (``oil'' and ``water''), under the mechanism of Rayleigh-Taylor instability. A  similar 3D DNS simulation has been performed by authors of  \cite{vincent} (and  previously, a 2D case had been described in  \cite{labourasse}). A parametric study for the same case has been published by Toutant  \cite{toutant}. These articles presented \textit{a priori} DNS studies\footnote{Using VOF method for interface tracking.} of magnitude, temporal evolution and parametric dependence of sub-grid tensors for LES, among which $\tau_{rnn}$ tensor which is the subject of our modelling efforts in this article. It is therefore essential to compare the results obtained by use of ADM-based $\tau_{rnn}$ model with mentioned works. 

\begin{figure}[ht!]
\centering
\includegraphics[scale=0.2]{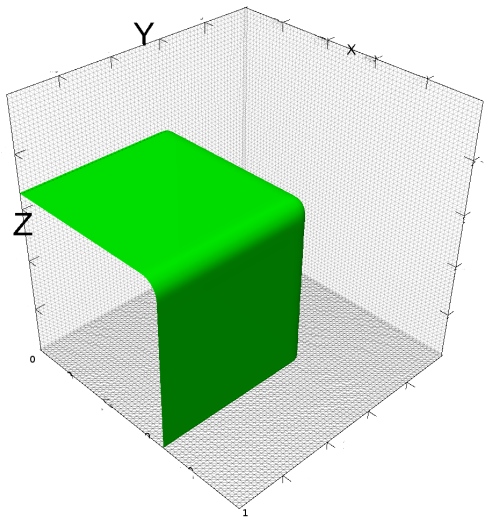}
\caption{Initial positioning of oil (green surface)  and water (rest of the domain) phases in a phase-reversion numerical simulation. Gravity is along vertical $Z$ axis. }\label{mix0}
\end{figure}

Basic description of the computational case is the following. In a cubical box of side length $L=1$ m , two phases are initially positioned in such way, that a lighter phase (oil), with $\rho_o=900$ kg.m$^{-3}$  occupies a cubic subset placed in octant closest to $(0,0,0)$ point. The other, heavier phase (water) with density of 1000 kg.m$^{-3}$ fills the rest of the domain. See Fig. \ref{mix0} for reference. Authors of  \cite{vincent} set $\left(\frac{1}{2}L\right)^3$ as oil cube dimension, whereas in our simulations it is most often $\left(\frac{2}{3}L\right)^3,$ which we motivate by our intention to yield greater interfacial surface on coarse grids which  cause poor mass conservation. It has been found that characteristics of $\tau_{rnn}$  -- such as its domain-averaged magnitude presented below -- do not change quantitatively due to that scaling. All physical parameters have been set identical to the reference work  \cite{vincent}, that is the viscosities of water and oil are respectively $0.001$ and $0.1$ Pa$\cdot$s, and surface tension is set to 0.075 N$\cdot$m$^{-1}$.

Three different grids have been used, namely $32\times32\times32=32^3$ grid ($32768$ nodes), the $64^3$ grid with $262144$ nodes, and the $96^3$ grid with 884736 nodes. Most of calculations were performed on $16$-processor clusters, containing $2210$MHz AMD Opteron processors, and $32$ GB of RAM. All simulations were prepared using SAILOR-LES solver, using the Smagorinsky model for $\tau_{luu},$ and CLSVOF module for interface tracking; GFM technique was applied to implicitly treat pressure and density jumps on the interface.

\begin{figure}[ht!]
\centering
\includegraphics[scale=0.4]{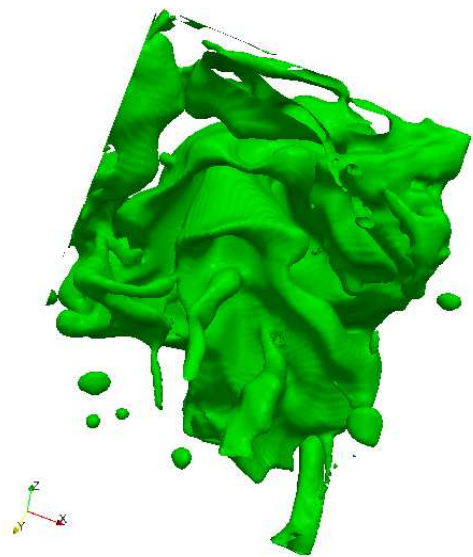}
\caption{Interface shape for $t=7.6$ obtained on the $96^3$ grid.}\label{96look}
\end{figure}

\begin{figure*}[ht!]
\centering
\includegraphics[scale=0.29]{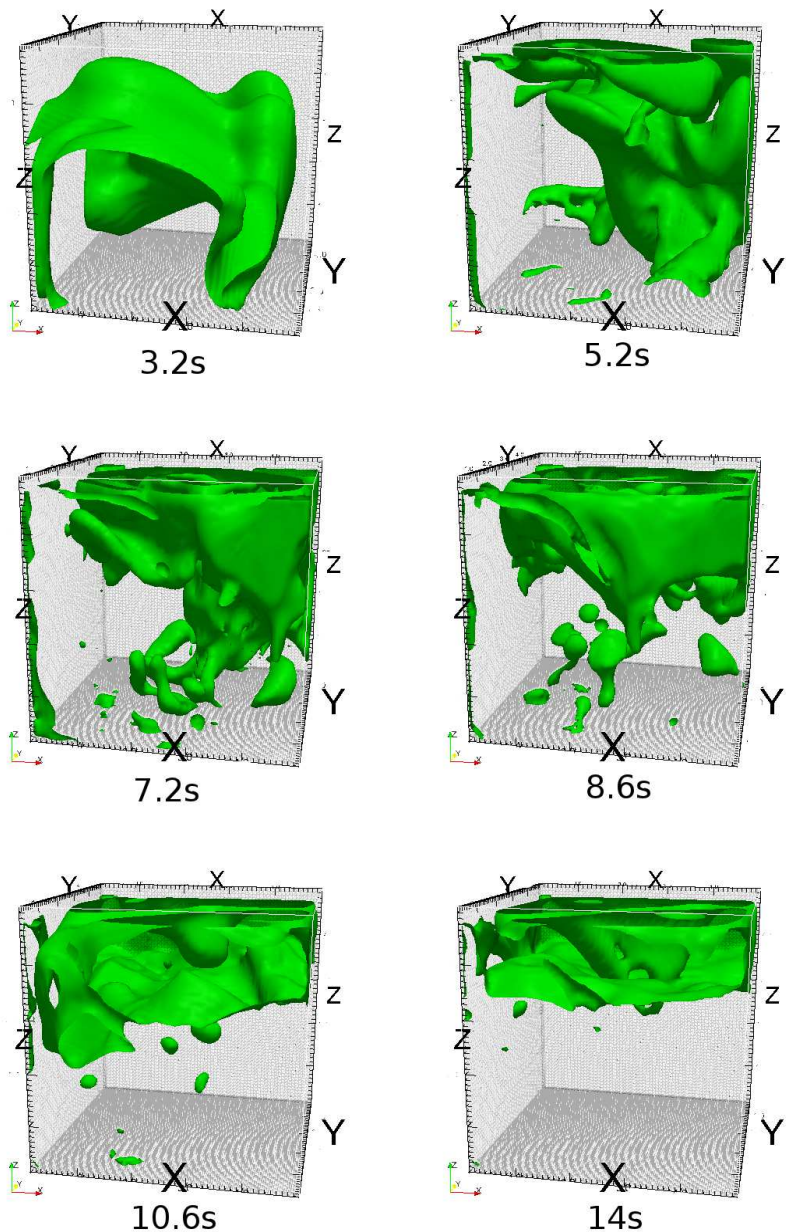}
\caption{Macroscopic evolution of the interface for oil-mixing test case using a  $64^3$ grid. }\label{64_evo}
\end{figure*}

\begin{figure}[ht!]
\centering
\includegraphics[scale=0.4]{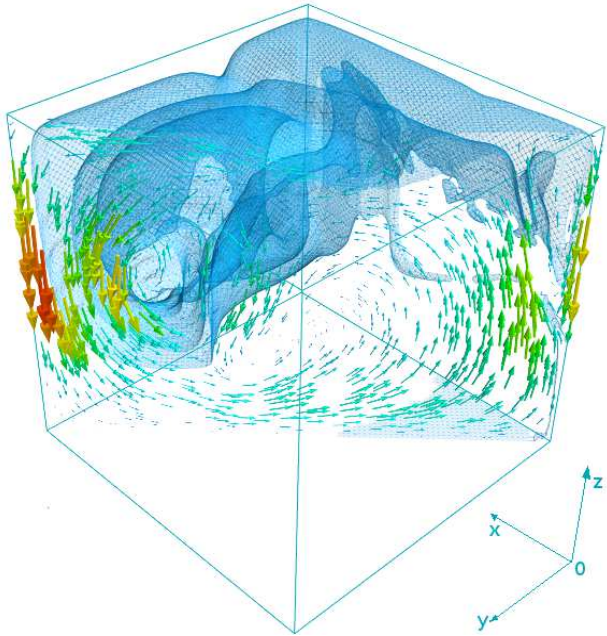}
\caption{Example of velocity vectors drawn along with wireframe interface representation.}\label{mix_wire}
\end{figure}

Figure \ref{64_evo} displays  temporal macroscopic evolution of the interface. Visible in the figure is the large-scale movement of lighter phase upwards, caused by buoyancy force, after which it hits opposite corner of the domain (also see Fig. \ref{mix_wire}), causing large vortical structures to emerge, which subsequently leads to creation of small-scale interfacial structures, such as droplets and ligaments,  best resolved on the finest grid. However, it is evident from inspection of Fig. \ref{64_evo} and \ref{96look}  that overall number of structures is smaller than described in DNS simulation  \cite{vincent} (obtained using $128^3$ grid and the VOF method). 

\begin{figure}[ht!]
\centering
\includegraphics[scale=0.45]{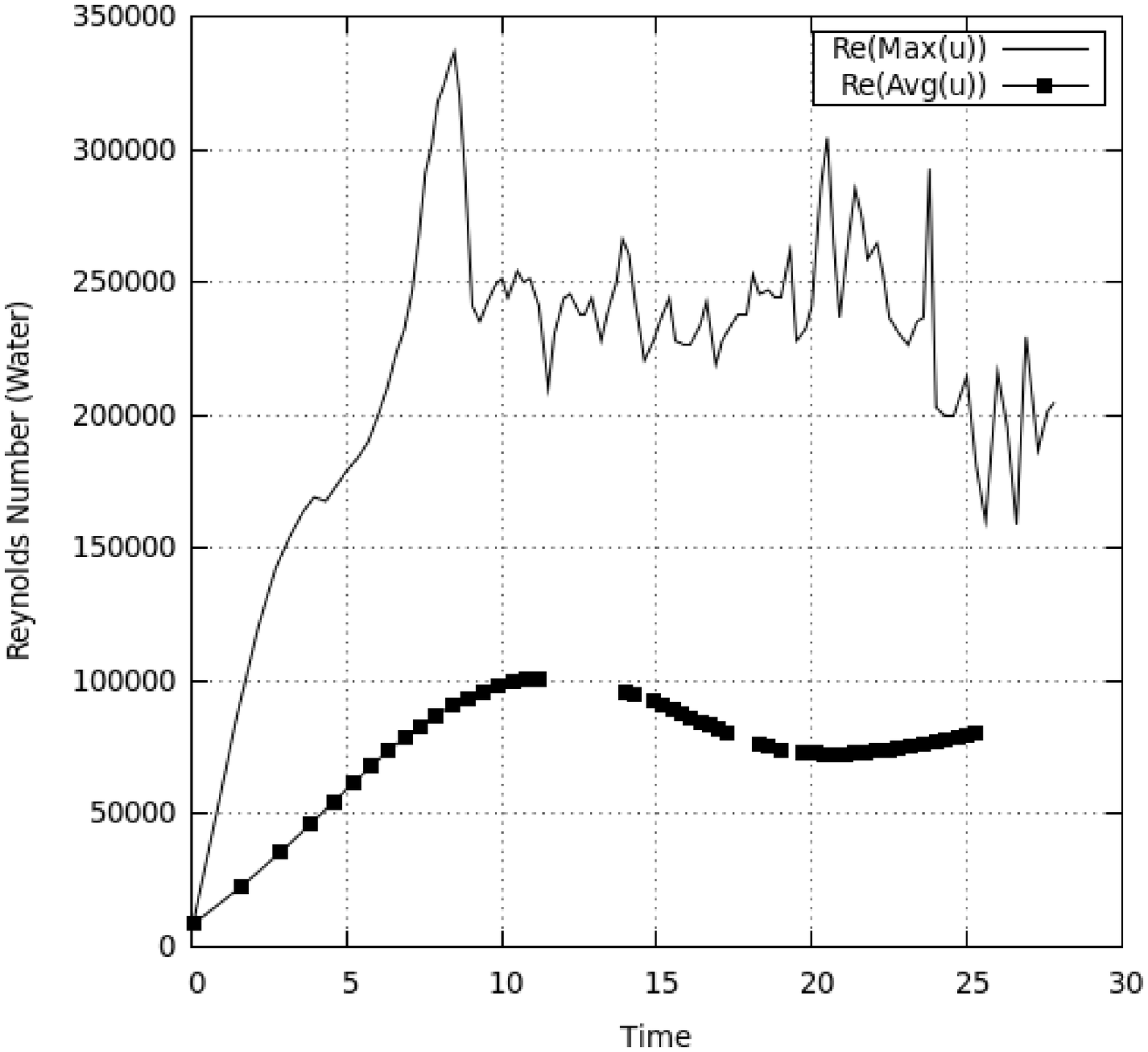}
\caption{Reynolds number in water temporal evolution. Time units: $1/5$ s.}\label{mix_re}
\end{figure}

Authors of  \cite{vincent,labourasse, toutant} use dimensionless number $Re, $ defined using domain size as characteristic dimension  and varying approaches towards characteristic velocity. Let us now review these approaches and apply them to presented computational case. Larocque et al.  \cite{larocque} defines $Re$ as 

\begin{equation}
\label{mix21}
Re=\frac{\rho_w L U_g}{2\mu_w}
\end{equation}
using ``predicted velocity'' $U_g$ defined with

\begin{equation}
\label{mix2}
U_g=\frac{\rho_w-\rho_o}{\rho_w}\sqrt{\frac{gL}{2}}.
\end{equation}
This time-independent quantity is therefore calculated using only species density and characteristic measure $L$ equal to the domain size  \cite{larocque}. For varying viscosity and surface tension coefficients,  \cite{larocque} describes $Re$ between $1110$ and $554000\rb$ in their simulation. In the discussed simulation, when calculating using (\ref{mix2}), we get value of $Re=72827.$ 

It is clear that $Re$ is time independent in such interpretation; in contrast, Vincent et al.  \cite{vincent} define Reynolds number using maximum of $\ub_w$ (in water) at every time-step, and water viscosity, resulting in values up to $60000.$ In our work, when using $\max \ub_w,$ very high $Re$ values are observed (up to $3.5\cdot10^5$). In Fig. \ref{mix_re}, $Re$ values are plotted obtained using $\max \ub_w$ (continuous line).  These Reynolds number values, much higher than described in  \cite{vincent},  are caused by different definition of characteristic scale $L$. In  \cite{vincent} time-dependent value of $L$ is used, described by authors by ``size of larger eddy structures'', equal to the macroscale of turbulence  \cite{vincent_private}.  Similarily to (\ref{mix21}), in preparation of Fig. \ref{mix_re} constant $L=1$ has been used.

The second (squared) curve in Fig. \ref{mix_re}, is $Re$ calculated using domain-averaged  $\langle\bar{\ub}\rangle$ magnitude, which yields $Re$ at $10^5$ level. Averaging is performed in a way similar to presented in  \cite{vincent}, that is
\begin{equation}
\label{d5}
\langle\Psi\rangle=\frac{1}{V}\int\limits_V \Psi dV
\end{equation}
where $\Psi$ is an arbitrary scalar field and $V$ is computational domain, discretisation of $\Omega.$

These levels of $Re$ below $10^5$ are in agreement with constant ``expected'' values yielded by formulas (\ref{mix21}) and (\ref{mix2}).

\begin{figure*}[ht!]
\centering
\includegraphics[scale=0.9]{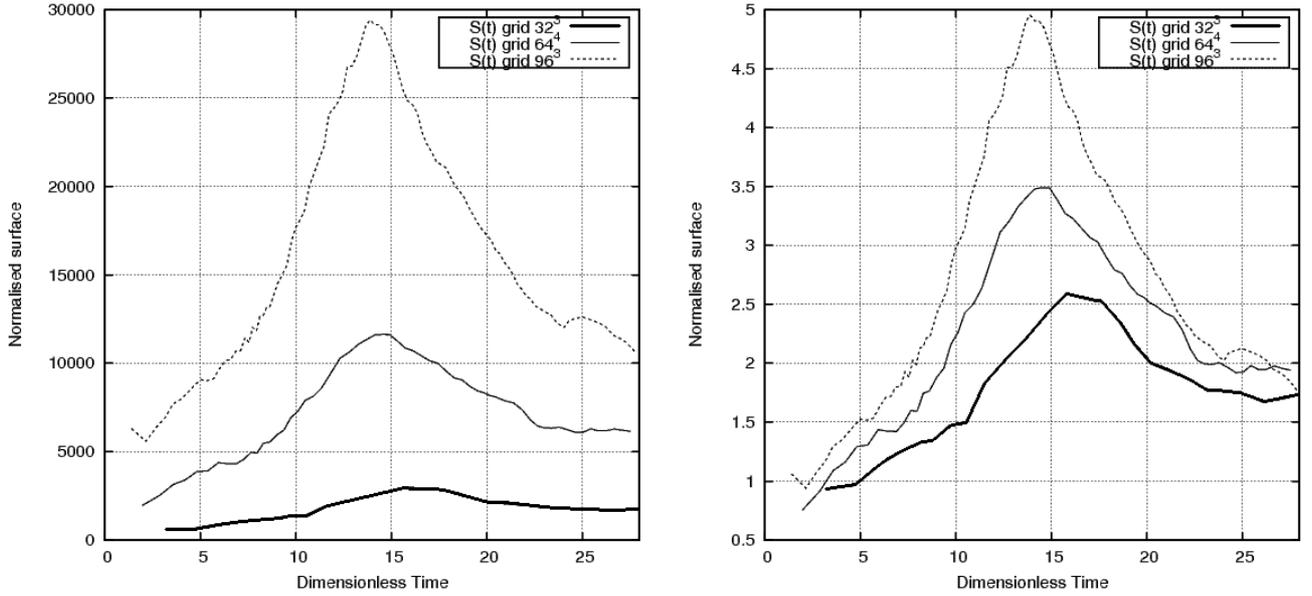}
\caption{Approximated interfacial area; temporal evolution for phase mixing case. Raw cell count (left) and values normalized by initial interfacial cell count (right). }\label{surf_plot}
\end{figure*}

Interfacial surface has been traced during our simulations, similarly to previous subsection, and the results are given in Figure \ref{surf_plot}. Interfacial area has been calculated using simple approximation described in  \cite{dropturb}, that is as a number of cells with $|\phi_{ijk}| \le d,$ where $d=\Delta x /4.$ In other words, this approach counts cells with values of level-set $\phi$ function small enough to assume they contain the interface. Such approach, however crude, is reminiscent of  \cite{vincent}, where authors propose a technique based on VOF $C$ function values. Neither of such approaches involves any geometrical reconstruction of the interface, and thus are equally dependent on discretisation and can be perceived only as rough approximations of surface area. Notwithstanding this possible lack of precision, we may observe that overall trend is present in Fig. \ref{surf_plot} of surface area rising to approximately $5$ times its original value on a $96^3$ grid, while on more coarse grids is characterized by less steep decrease following peak values. In particular we see that highest interface fragmentation took place at about $15$ time units. Below, data presented in Fig. \ref{surf_plot} will allow us to seek correlations between surface area and values of sub-grid surface tension tensor.

Note that for different parameters such as smaller ``oil'' phase  density (or greater ``water'' density), kinetic energy resulting from buoyancy will be inversely proportional to $\rho_o$ and thus cause interfacial surface to be greater by causing stronger oil fragmentation. The same behaviour, but pertaining to interfacial sub-grid term $\tau_{rnn}$ is visible in parametric DNS study of Larocque  \cite{larocque}. 

\begin{figure*}[ht!]
\centering
\includegraphics[scale=0.18]{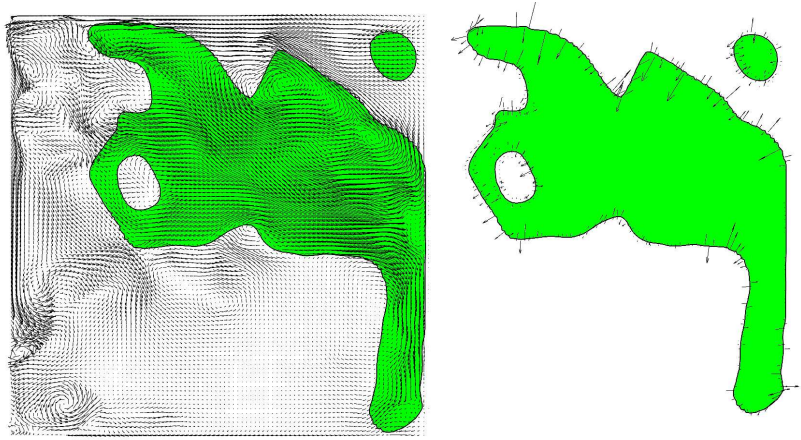}
\caption{Two cut-planes $y=2.5$ with vectors of velocity (left, vector magnitude reduced $0.4$ times) and $\tau_{rnn}$ (right, vectors enlarged $10$ times). Both snapshots were taken for $t\approx 20.$ Are inside ``oil'' phase is colored green.}\label{cactus}
\end{figure*}

Examples of velocity field visualizations for a $96^3$ grid are visible in Fig. \ref{cactus} (leftmost image) and Fig. \ref{mix_wire}. Latter figure displays the scale of vortical motion in the most kinetic part of simulation, right after the overturned oil mass hits opposite domain corner. In the figure, only half of the interfacial surface is visible as a wireframe. Vectors of velocity in \ref{mix_wire} are rescaled accordingly to their magnitude, while in Fig. \ref{cactus} similar scaling has been performed by much smaller coefficient, to produce clear image. 

Figure \ref{mix_wire} contains velocity cut-plane parallel to $z$ axis which contains $x=y$ line. The visible part of the interface lies in $y<x$ part of the domain. 

In Fig. \ref{cactus}, a $y=2.5$ cut-plane has been drawn. Apart from representing velocity vector field, Fig. \ref{cactus} contains (rightmost image) graphic representation of $\tau_{rnn}$ vector field, taken at the same moment of dimensionless time. 

\begin{figure}[ht!]
\centering
\includegraphics[scale=0.9]{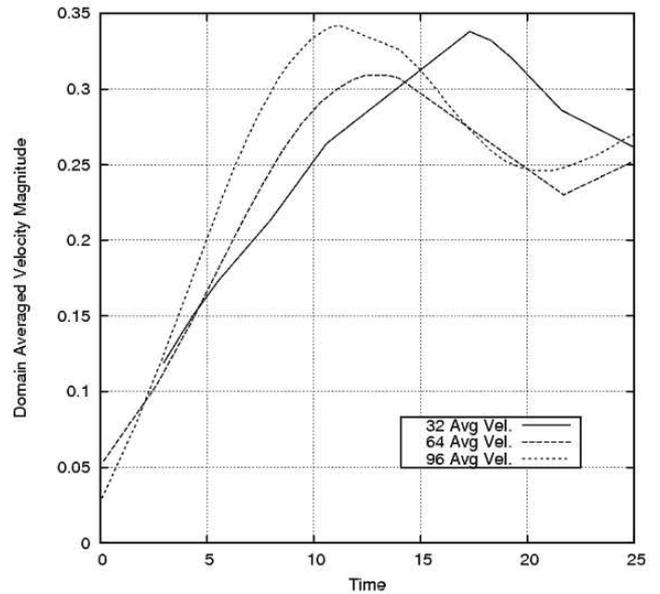}
\caption{Domain-averaged velocity (m/s)  distribution in discretisation dependence in $\rho_o=900$ kg.m$^{-3}$ case.}\label{mix_avg_vel}
\end{figure}

Velocity field characteristic was presented in Fig. \ref{mix_re}; to further characterize resolved $\bar{\ub}$ field we additionally present the plot of the domain-averaged velocity distribution $\langle \bar \ub \rangle$ for three considered grids in Fig \ref{mix_avg_vel}. During first $10$ time units, acceleration due to overturning motion is clearly visible in all cases. Subsequently the velocity drops, as oil and water masses begin quasi-periodic  \cite{vincent} sloshing movements  during which any remaining kinetic energy is dissipated by viscous effects. 

\begin{figure*}[ht!]
\centering
\includegraphics[scale=0.85]{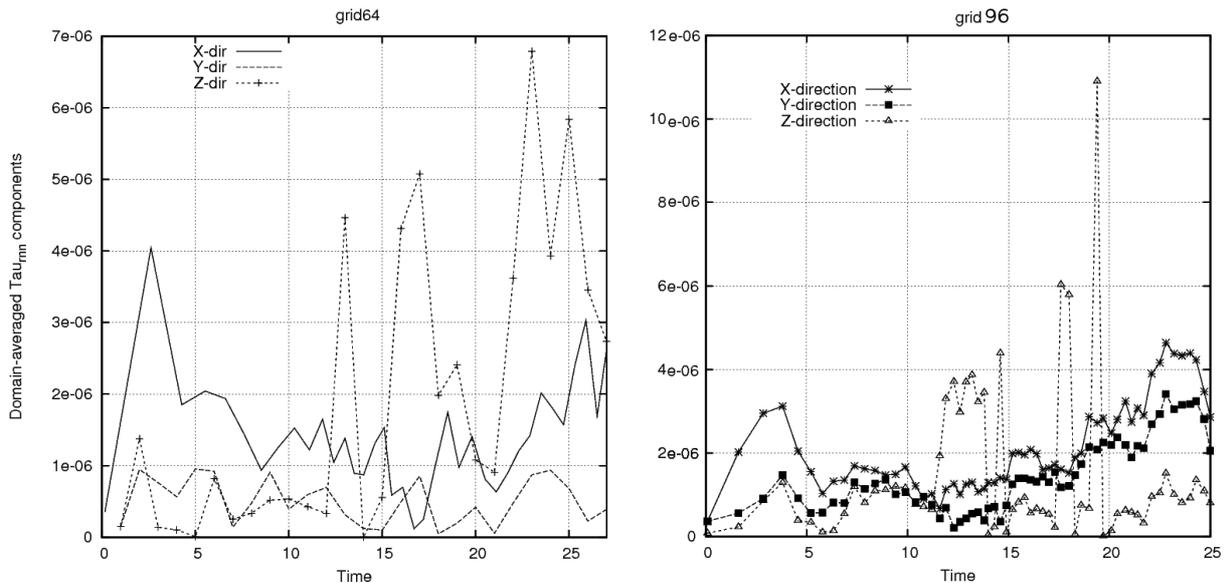}
\caption{Temporal evolution of $\langle\tau_{rnn}|_{x,y,z}\rangle$ components, no scaling.}\label{avgcomp}
\end{figure*}

In Fig. \ref{avgcomp} non-rescaled, averaged components $x,y$ and $z$ of $\tau_{rnn}$ tensor are plotted. Visible are curves for $64^3$ and $96^3$ grids, for $t\in \lb 0,25 \rb.$  As the curves for $\langle \tau_{rnn}|_x\rangle$ and $\langle \tau_{rnn}|_y \rangle$ show, the $x$ and $y$ components do not rapidly increase during the simulation, while $\langle \tau_{rnn}|_z\rangle$ forms peaks visible in interval $\lb 10,25\rb$ of both plots. Reason for this is linked to dominating character of buoyancy force in simulation, as overturned mass of oil ruptures into droplets. DNS simulations show great fragmentation in this phase  \cite{larocque} followed by dispersed flow which was not captured in present LES simulations, due to mass loss and numerical coalescence; however medium-sized and large interfacial formations are captured. It is to these formations that peak $\tau_{rnn}|_z$ values seem linked, as concluded by Vincent in  \cite{vincent}. Still, in  parts of the simulation where $z$-direction peaks do not occur, all components act similarly, and overall averaged norm of $\tau_{rnn}$ remains at constant level, a behaviour which we will describe below. Large peak of $\langle \tau_{rnn}|_x\rangle$ curve in $t\in\lb 0,5\rb$ interval visible in both plots of Fig. \ref{avgcomp} may be connected to movement of oil drop towards opposite domain corner at the onset of simulation. 

\begin{figure*}[ht!]
\centering
\includegraphics[scale=0.3]{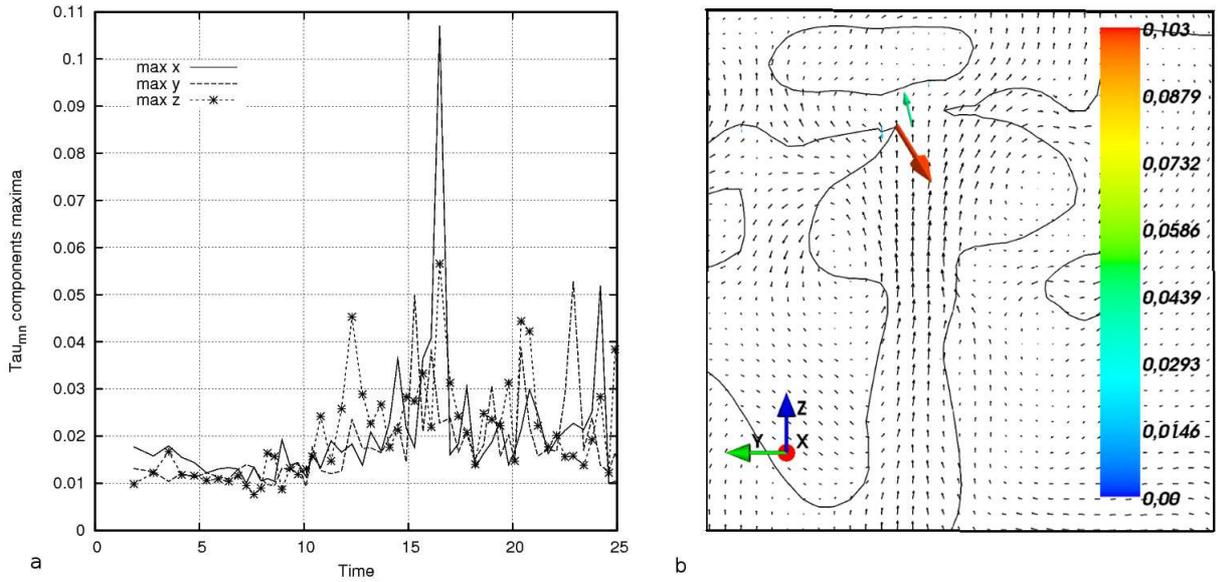}
\caption{(a):Temporal evolution of $\max(\tau_{rnn}|x),\max(\tau_{rnn}|y)$ and $\max(\tau_{rnn}|z)$ for oil mixing simulation. No averaging was performed. (b): Cutplane drawing for $t=16.4,$ matching peak value in (a). Red vector (unscaled) represents peak $\tau_{rnn}$ vector, also visible is 2D $zy$ cutplane of the interface (thick black line), and velocity field visualization (downscaled black vectors).}\label{maxima}
\end{figure*}

In Figure \ref{maxima}a, the maximum values of $\tau_{rnn}$ tensor are presented. The graph has been prepared for simulation on $64^3$ grid, using $\rho_o\colon \rho_w=0.7\colon 1$ ratio, which the results in more emphasized overturning motion and more interface defragmentation. Contrary to what may have been expected from averaged values, maximum levels are of order $10^{-1}$. Highest values ($x$-component peak at $t=16.4$) are once again connected with interfacial rupture phase in $t\in \lb 15,20 \rb$ interval, during which the mass divides into bulk part that forms large ligament that later ruptures into droplets. We have traced spatial location of peak tensor value, which is shown in Fig. \ref{maxima}b as being connected with interface cusp.

Concerning surface tension coefficient $\sigma$, it is important to remind that $\sigma$ is present in definition (\ref{lst2}) of $\tau_{rnn}$ and any change to $\sigma$ will cause $\tau_{rnn}$ to directly re-scale. However, some characteristics of the flow in the discussed case would also change then, e.g. greater surface tension will produce different, less fragmented oil mass configuration, also influencing the  average velocity of the overturning motion. At the same time, $\tau_{rnn}$ will become larger due to scaling by $\sigma.$ Because of that and other reasons -- such as the need to assess importance of two-phase specific terms altogether -- ratio of $\tau_{rnn}$ components might be calculated versus some flow-specific quantity, as mean velocity or flow inertia. Such ratio was used in  \cite{vincent}. 

In our work, we have assessed the significance of $\tau_{rnn}$ by comparing its norm  with that of the domain-averaged resolved inertial term 

\begin{equation}
\label{irt}
\rho  \overline{\ub}\otimes\overline{\ub},
\end{equation}
computed directly in simulation. The ratio of \[\frac{\tau_{rnn}}{\rho\overline{\ub}\otimes\overline{\ub}}\] is a dimensionless quantity, directly comparable with plots presented in reference works  \cite{vincent} and  \cite{labourasse}. Moreover, it may be seen as a comparison between tensors $\tau_{rnn}$ and $\tau_{luu}$ (that is the ``classical'' SGS tensor (\ref{ff4})) . When making the plots, we have used either the ratio 

\begin{equation}
\label{ratio1}
\frac{\langle | \tau_{rnn} | \rangle}{\langle || \rho  \overline{\ub}\otimes\overline{\ub} || \rangle },
\end{equation}
of tensor norms, or ratios using components  of sub-grid surface tension tensor such as
\begin{equation}
\label{ratio2}
\frac{\langle \tau_{rnn}|_z \rangle}{\langle || \rho  \overline{\ub}\otimes\overline{\ub} || \rangle }.
\end{equation}
Averaging operator $\langle \cdot \rangle$ was described in previous subsection. In the above equations, we have used Euclidean norm (Frobenius norm) of the matrix in the denominator, namely, for an arbitrary matrix $\mathbf{A}$:

\begin{equation}
||\mathbf{A}||=\sqrt{\sum\limits_i\sum\limits_j |a_{ij}|^2}.
\end{equation}

\begin{figure}[ht!]
\centering
\includegraphics[scale=0.8]{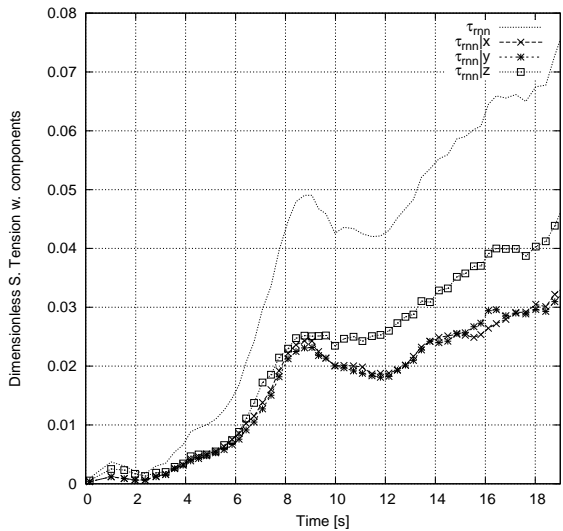}
\caption{Average normalized magnitude of $\tau_{rnn}$ tensor and its components, non-dimensional, concerning the simulation  on a $32^3$ grid.}\label{new32}
\end{figure}

In Figure \ref{new32}, an actual time dependence of fractions (\ref{ratio1}) and (\ref{ratio2}) can be seen.  Figure \ref{new32} has been prepared using least refined grid $32^3,$ and indicates a constant rise in plotted value, reaching $7\%$ of approximated inertia for $t=16$ s. This is comparable to values described in  \cite{vincent}, in which the authors give  e.g. the value of (\ref{ratio2}) at about 
$5\%$ level for $t=15s.$ Also, Figure (\ref{new32}) shows the values of individual tensor components in the same simulation - the domination of the $z$ component (due to buoyancy force generating the flow) is well pronounced, which again reflects the \textit{a priori} DNS results.  It  shows that ADM-based reconstruction approach to $\tau_{rnn}$ tensor allows for sensible reconstuction of the sub-grid surface tension force, whose magnitude measured with respect to (\ref{irt}) raises as the flow moves to the viscous dissipation phase.

\begin{figure}[ht!]
\centering
\includegraphics[scale=0.8]{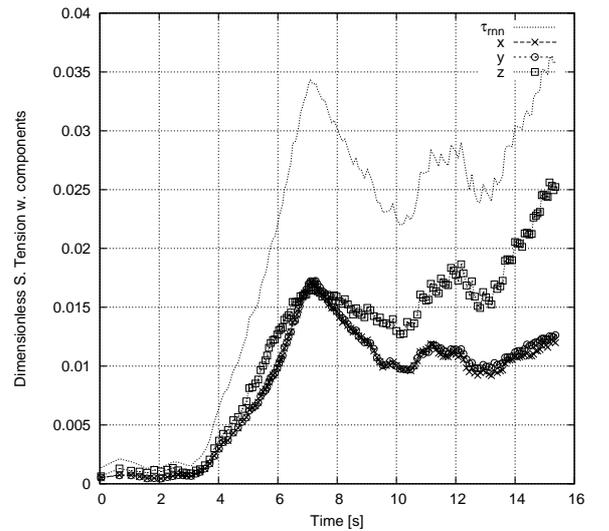}
\caption{Average normalized magnitude of $\tau_{rnn}$ tensor and its components, non-dimensional, concerning the simulation  on a $64^3$ grid.}\label{new64}
\end{figure}

\begin{figure}[ht!]
\centering
\includegraphics[scale=0.2]{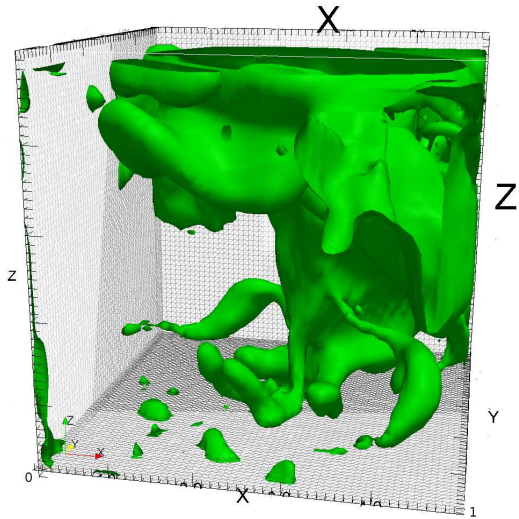}
\caption{Interfacial surface for $t\approx 7$s in a $64^3$ simulation, corresponding to peak value in Fig. \ref{new64}.}\label{64peak}
\end{figure}

\begin{figure}[ht!]
\centering
\includegraphics[scale=0.21]{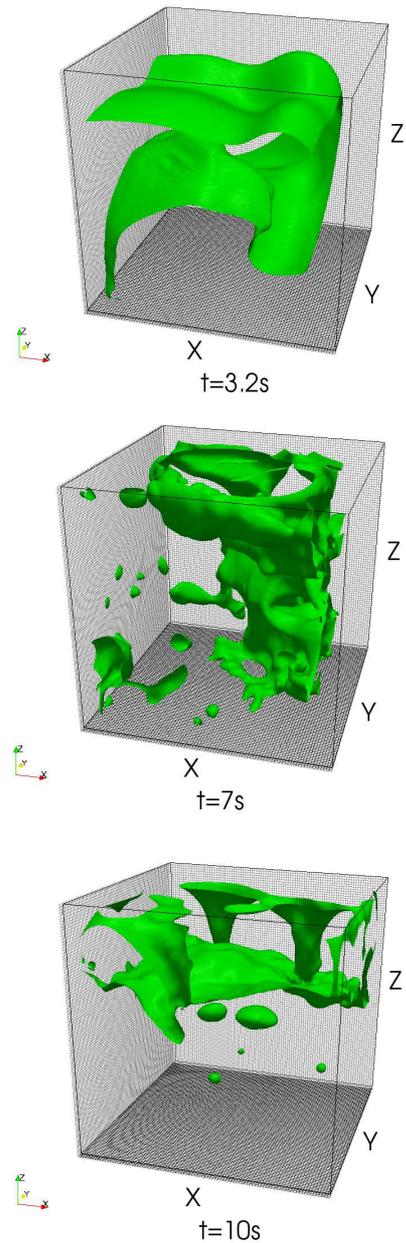}
\caption{A $96^3$ simulation - macroscopic evolution of the interface without the inclusion of $\tau_{rnn}$ tensor.}\label{96nt}
\end{figure}

\begin{figure}[ht!]
\centering
\includegraphics[scale=0.95]{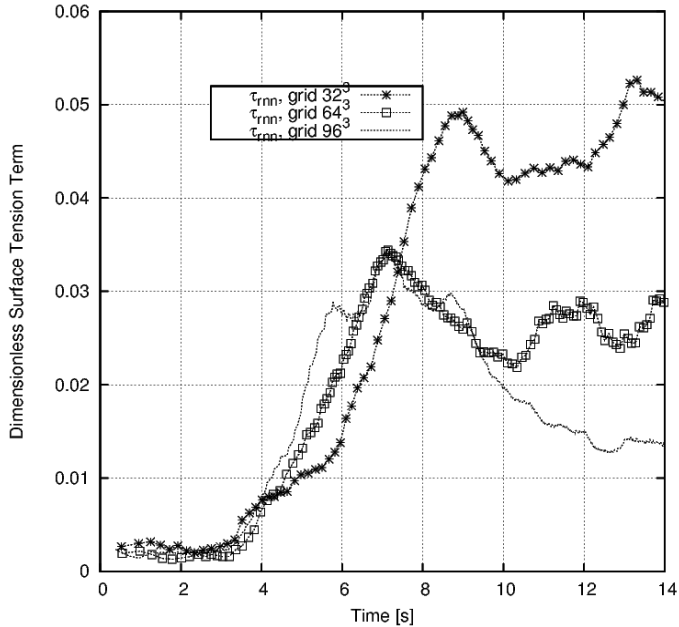}
\caption{Average normalized magnitude of $\tau_{rnn}$ tensor and its components, non-dimensional, concerning the simulation  on a $96^3$ grid, plotted together with similar curves for $32^3$ and $64^3$. }\label{96-6s}
\end{figure}

As we can observe in Figure \ref{new64}, the behavior of the model is similar when a denser, $64^3$ node grid is used; we observe a raising value of ratio (\ref{ratio1}) and the shape of the curve is similar to Fig \ref{new32} with a pronounced peak for $t\approx 7s.$ This peak is linked to a large value of the interfacial surface, as can be seen in Fig. \ref{64peak}. Overall values of plotted ratio for $\tau_{rnn}$ tensor  are about $60\%$ of values for $32^3$ grid (Figure \ref{new32} and \ref{96-6s}), which could be attributed to decrease of difference between $\ub$ and $\ub^*$ velocity fields (and subsequently between the filtered and reconstructed normal vector fields) due to the fact that the flow is less under-resolved.   The level of $\tau_{rnn}|_z/||\rho(\ub\otimes\ub)||$ is comparable to what is indicated in the DNS \textit{a priori} test, and also its domination over other components is in full agreement with published work of Vincent  \cite{vincent}; it becomes even more apparent towards the end of the plotted time interval. To offer more insight into this particular simulation using $64^3$ grid, we attach a more detailed view of concurrent interface shapes, visible in Figure \ref{64_evo}.

Finally, similar in character and observed magnitudes is the evolution of the $\tau_{rnn}$ tensor when a $96^3$ grid is used, as can be observed in Figure \ref{96-6s}. In the Figure, magnitudes from previous plots have been placed to enable a comparison. As can be seen in Fig. \ref{96-6s},  although the most refined grid yields similar behaviour in first six seconds of simulation, the averaged magnitude of the surface tensor drops in subsequent stages even more rapidly than for $64^4$ case. This could support the conclusion about the velocity field being less under-resolved on denser grids. Notwithstanding this facts, it has to be noted that  \cite{vincent} describes different tendency in the plots of (\ref{ratio2}) and of other $\tau_{rnn}$ components, that is, magnitudes of these ratios are increasing proportionally to the number of grid points.

\begin{figure}[ht!]
\centering
\includegraphics[scale=0.3]{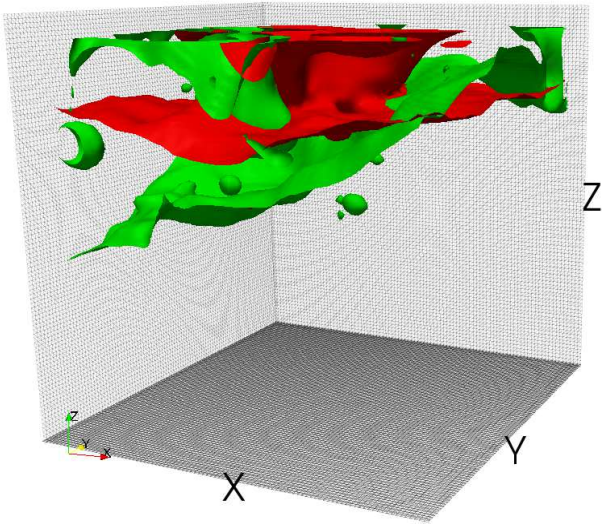}
\caption{Simulation using the $96^3$ grid with (green surface) and without (red) the inclusion of the $\tau_{rnn}$ tensor; $t\approx 11s$. }\label{96comp1}
\end{figure}

An example of the interface shape obtained for $t=7.6s$ using a $96^3$ grid is presented in Figure \ref{96look}, displaying a number of resolved film and ``finger'' formations. Additionally, simulation on a finest grid performed \textit{without} the use of the ADM-$\tau$ model is presented in Figure \ref{96nt}, wich is to certain extent comparable with Figure \ref{96look}. However, subsequent of the interface evolution even on this finest grid changes with the inclusion of ADM-$\tau$ model, as presented further in Figure \ref{96comp1}.

We conclude by remarking that inclusion of $\tau_{rnn}$ force into the simulation of oil--water mixing case results in visible macroscopic differences (Figures \ref{96comp1} and \ref{disfigure3}) between obtained interfacial shapes. This is expected since $\tau_{rnn}$ exerts small, but constant influence  on the interfacial geometry over entire simulation. Although overall mechanism of mixing with and without model usage  is very similar, some ligaments, droplets and fluid ``fingers'' are either shifted in position or size, or not present at all. This difference can be observed in Fig. \ref{96comp1}, in which the inclusion of $\tau_{rnn}$ is shown to have caused a shift in the bulk interface position and creation of more small-scale features. Overall assessment of this result - in other way than comparison with DNS \textit{a priori} simulation would require experimental data, or DNS results of simulations using very fine discretisation to assure that simulation is fully resolved. Meanwhile authors of  \cite{vincent} conclude in their paper that for DNS to be resolved in this case, a grid of at least $128^3$ is still not sufficient.

\subsection{Further Assessment of $\tau_{rnn}$ Inclusion}

\begin{figure}[ht!]
\centering
\includegraphics[scale=0.165]{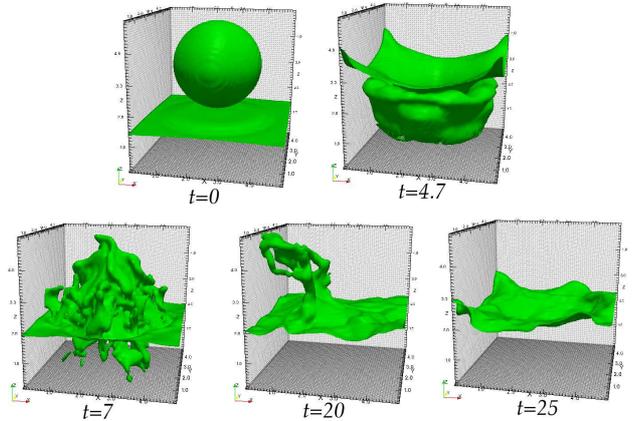}
\caption{Macroscopic definition for the ``splash case''.}\label{disfigure1}
\end{figure}

As a means to assess the influence of the $\tau_{rnn}$ inclusion into the simulation, consider the case of a large droplet of heavy fluid in free fall, embedded in lighter fluid; similar to water droplet in air, although density ratio has been set to $\rho_w\colon\rho_a=10\colon 1$ for stability reasons. In Figure \ref{disfigure1} the macroscopic, temporal evolution of such simulation, computed on a $64^3$ grid, is presented.

As it can be seen if Fig. \ref{disfigure1}, the droplet enters quiescent fluid mass after which a waving motion begins with bulk fluid mass oscillating in a way resembling a membrane (e.g. between $t=4.7$ and $t=7$ in Fig. \ref{disfigure1}).  For $t=7$ and $t=20,$ two of the uppermost positions of the oscillating mass are visible, when top wall of the domain is reached by it. Final image presented in Fig. \ref{disfigure1} for $t=25$ presents a more quiescent fluid surface. 

\begin{figure}[ht!]
\centering
\includegraphics[scale=0.42]{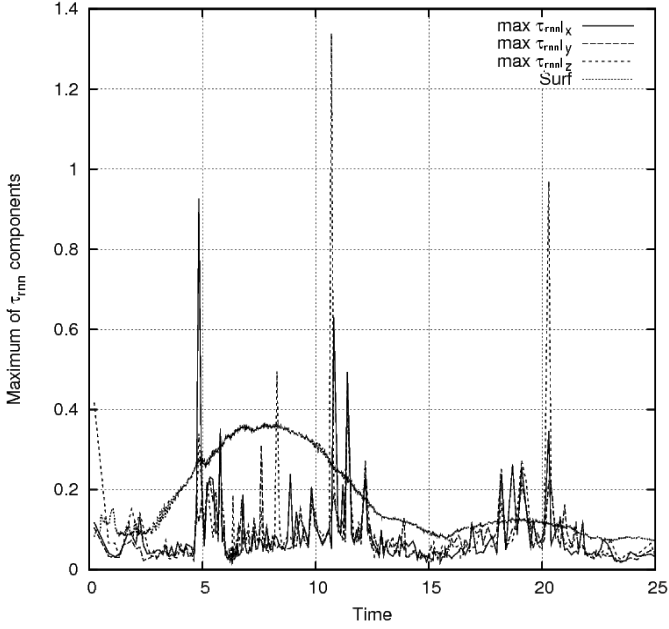}
\caption{The $\tau_{rnn}$ tensor components maxima, plotted together with rescaled surfacial area for the splash case.}\label{disfigure2}
\end{figure}

For this particular simulated case, it is easy to observe that maximum values of $\tau_{rnn}$ tensor maxima is correlated with aforementioned ``uppermost'' positions of the oscillating mass, when surfacial area also has its peaks. This is easily observable in Figure \ref{disfigure2}, where the maxima have been plotted together with rescaled surface area of the interface. Virtually all of the peak values of $\tau_{rnn}|_z$ curve occur in $t\in\lb 4,12\rb$ and $t\in \lb 16,22\rb$ intervals, that correspond to ``uppermost'' positions of bulk mass during the first oscillations\footnote{We do not claim that the occurence of $\tau_{rnn}$ maxima implies high level of ratio (\ref{ratio1}). Maximum may occure locally - even in a single grid cell - and be traced to a specific interfacial formation as shown in Fig. \ref{maxima}.}.

\begin{figure}[ht!]
\centering
\includegraphics[scale=0.25]{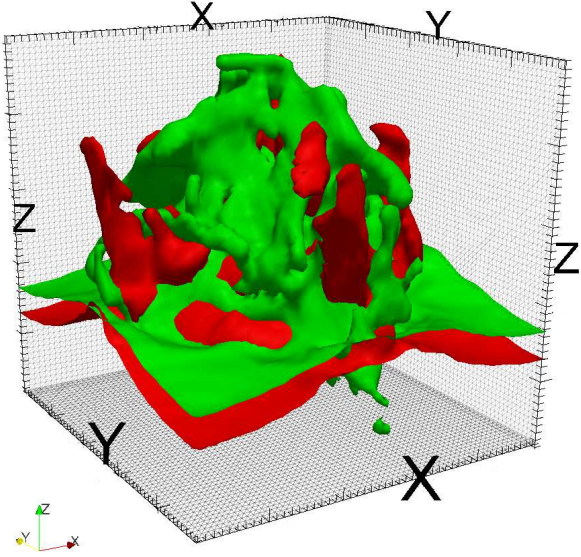}
\caption{Comparison of the interface shapes in droplet splash simulation. Green surface corresponds to simulation including $\tau_{rnn}$ tensor. Red surface corresponds to simulation in which the tensor was omitted.}\label{disfigure3}
\end{figure}

\begin{figure}[ht!]
\centering
\includegraphics[scale=0.9]{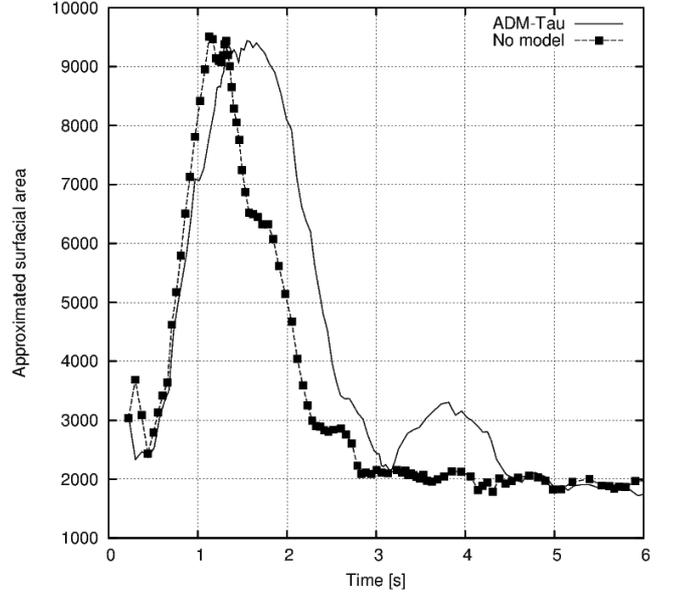}
\caption{Approximated surfacial areas for simulations with (continuous line) and without the use of the ADM-$\tau$ model, for the splash case.}\label{disfigure4}
\end{figure}

Such behavior of the tensor maxima suggests that $\tau_{rnn}$ tensor is not random, instead it is strongly dependent both on resolved interfacial shape and interfacial area. This could falsify any possible hypothesis saying that our results are merely a generation of a random vector field. Moreover, inspection of Figure \ref{disfigure3} convinces us that the influence of tensor in equation (\ref{ns+tau}) is significant -- simulations with and without $\tau_{rnn}$ are divergent even in this preliminary stage, which is in accordance with conclusions of Vincent  \cite{vincent} that influence of this tensor will be significant even if no small-scale fluid formations are yet present. For the same simulation, a comparison of temporal realisations of approximated interfacial areas is shown in Fig. \ref{disfigure4}. Apart from the discernible difference in peak shapes for $t\in\lb 1,2\rb$, an additional peak is visible in the simulation including ADM-$\tau$ model. 

\section{Discussion}
We have presented and tested an approach of reconstructing sub-grid surface tension force in LES simulation, thus showing that introduction of one of LES-specific tensors emerging in two-phase flow is feasible with reasonable computational cost.  Results of sample calculations indicate that values and behaviour of $\tau_{rnn}$ force are similar to DNS-based \cite{vincent}. However, the ``Sub-Grid Mass Transfer'' (SMT) scalar term $\sigma_1$ discussed in Section \ref{tau_section} is not modelled yet, therefore we do not claim to have achieved a complete LES of two-phase flow. Technically, to cite Gorokhovski \& Herrmann \cite{gorokhovski}, the presented technique still has to be viewed as a  \textit{quasi-LES/DNS} simulation, since the SMT term is not modelled. 

Evolution of $\tau_{rnn}$ sub-grid surface tension force term presented in this work is based on its definition, and rests more on a mathematically justified deconvolution of $\ub^*$ by ``inverted filtration'' rather than being anyhow rooted in flow physics, e.g. by forming a relation between primitive variables and resulting $\tau_{rnn}.$ By this, we mean that ADM-based $\tau_{rnn}$ calculation is not a ``model'' \textit{par excellence}, except in one sense: an assumption that $\nb^*=\Psi(\ub^*)$ operation described in Sect. \ref{tau_section} is a justified reconstruction of ``deconvoluted curvature'' $\kappa^*$. The latter quantity is meaningful in that $\tau_{rnn}$ is essentially a force created by the difference between resolved (filtered) and unresolved (subfilter) curvatures $\overline{\kappa}$ and $\kappa$. Present authors are however aware of potential drawbacks in presented model. For this reason, further studies could seem valuable in this field, such as:

\begin{itemize}
\item An ever more comprehensive parametric study of ADM-$\tau$ method to further prove its correctness with respect to DNS \textit{a priori} calculations, similar to  \cite{larocque};
\item Comparison of ADM-$\tau$  results with different possible approaches, such as direct reconstruction of $\nb^*$ field or $\kappa^*$ deconvoluted curvature. Basing on methodology introduced in this article, the simplest method that yields correct results should be continuedly developed;
\item Performing simulations for which more experimental data exist, such as using ADM-$\tau$ method for simulating atomization. Characteristics of those processes, like droplet distribution histograms, could be affected by $\tau_{rnn}$ force;
\item Further work could also include modelling of $\sigma_1$ ``sub-grid mass transfer''  tensor.
\item Unlike direct reconstruction of $\nb^*$ or $\kappa^*,$ the ADM-$\tau$ scheme presented in this work is suitable for use with Domaradzki's reconstruction scheme  \cite{domaradzki} for $\ub^*.$ Hence calculations using it should constitute a good possibility for testing the method. Besides, in its original implementation, the Domaradzki scheme includes usage of dense grid, which could be very useful in calculating aforementioned ``sub-grid mass transfer'' tensor directly.
\end{itemize}

\section{Finishing Remarks}

The ADM-$\tau$ model has been tested and its results compared with \textit{a priori} DNS results yielding satisfactory results, such as the temporal evolution of the ratio of norms of the ADM-reconstructed $\tau_{rnn}$ tensor and resolved inertia. 

We believe this study should be followed by propositions of models for sub-grid mass transfer models, and assessments of the performance of ADM-$\tau$ model in other computational cases, at least to the extend permitted by availability of the DNS and experimental data.

\section{Acknowledgements}
Parts of this work were performed within the TIMECOP -- AE (Toward Innovative Methods for Combustion Prediction in Aero-Engines) Project, co-funded by the European Commission within the Sixth Framework Program. Project no: AST5-CT-2006-030828. 

The visualizations have been prepared using Mayavi  \cite{mayavi}, Paraview  \cite{kenneth} and Gnuplot  \cite{gnuplot}.

We would like to thank the anonymous reviewers of the article for valuable questions and suggestions.

\section*{List of Acronyms}
\begin{itemize}
\item VOF - Volume of Fluid Method
\item LS - Level Set Method
\item LES - Large Eddy Simulation 
\item CLSVOF - Coupled Level Set-Volume of Fluid method
\item DNS - Direct Numerical Simulation
\item SMT - ``Subgrid Mass Transport''
\item GFM - Ghost Fluid Method
\item SAILOR - Spectral and High Order Low Mach-Number LES (flow solver)
\item ENO - Essentially Non-Oscillatory (differential scheme)
\item WENO - Weighted Essentially Non-Oscillatory (differential scheme)
\item PLIC - Piecewise Linear Interface Calculation
\item ADM - Approximate Deconvolution Method
\end{itemize}

\bibliography{wa_sst}
\end{document}